\documentclass{pasa}%

\title[Cosmological PAH surveys by SPICA]{Unbiased large spectroscopic surveys of galaxies selected by SPICA using dust bands}
\author[Kaneda et al.]{H. Kaneda$^1$\thanks{kaneda@u.phys.nagoya-u.ac.jp}, 
D. Ishihara$^1$, S. Oyabu$^1$, M. Yamagishi$^2$, T. Wada$^2$, 
L. Armus$^3$, M. Baes$^{4}$, 
V. Charmandaris$^5$, B. Czerny$^{6,7}$,
A. Efstathiou$^8$,
J. A. Fern\'andez-Ontiveros$^{9,10,11}$,
A. Ferrara$^{12,13}$,
E. Gonz\'alez-Alfonso$^{14}$, M. Griffin$^{15}$,
C. Gruppioni$^{16}$, E. Hatziminaoglou$^{17}$,
M. Imanishi$^{18,19,20}$, K. Kohno$^{21}$, J. Kwon$^2$,
T. Nakagawa$^2$, T. Onaka$^{22}$, F. Pozzi$^{23}$, D. Scott$^{24}$,
J.-D. T. Smith$^{25}$, L. Spinoglio$^9$, T. Suzuki$^1$, F. van der
Tak$^{26, 27}$, M. Vaccari$^{28, 29}$, C. Vignali$^{30, 31}$, \and L. Wang$^{26, 27}$\\
\affil{$^1$Graduate School of Science, Nagoya University, Furo-cho, Chikusa-ku, Nagoya 464-8602, Japan}%
\affil{$^2$Institute of Space and Astronautical Science, Japan Aerospace Exploration Agency, Chuo-ku, Sagamihara 252-5210, Japan}
\affil{$^3$Infrared Processing and Analysis Center, MS 100-22,
California Institute of Technology, Pasadena, CA 91125, USA}
\affil{$^{4}$Sterrenkundig Observatorium, Universiteit Gent, Krijgslaan 281 S9, B-9000 Gent, Belgium}
\affil{$^5$Institute for Astronomy, Astrophysics, Space Applications \&
  Remote Sensing, National Observatory of Athens, GR-15236, Penteli, Greece}
\affil{$^6$Center for Theoretical Physics, Polish Academy of Sciences, Al. Lotnik\'{o}w 32/46, 02-668 Warsaw, Poland}
\affil{$^7$Copernicus Astronomical Center, Polish Academy of Sciences, Bartycka 18, 00-716 Warsaw, Poland}
\affil{$^8$School of Sciences, European University Cyprus, Diogenes
Street, Engomi, 1516 Nicosia, Cyprus}
\affil{$^{9}$Istituto di Astrofisica e Planetologia Spaziali, INAF, Via Fosso del Cavaliere 100, I--00133 Roma, Italy}
\affil{$^{10}$Universidad de La Laguna (ULL), Dept. de Astrof\'isica, C/Astrof\'isico Fco. S\'anchez s/n, E--38206 La Laguna, Spain}
\affil{$^{11}$Instituto de Astrof\'isica de Canarias (IAC), C/V\'ia L\'actea s/n, E--38205 La Laguna, Spain}
\affil{$^{12}$Scuola Normale Superiore, Piazza dei Cavalieri 7, I-56126 Pisa, Italy}
\affil{$^{13}$Kavli IPMU, WPI, The University of Tokyo, Kashiwa, Chiba 277-8583, Japan}
\affil{$^{14}$Universidad de Alcal\'a, Departamento de F\'isica y Matem\'aticas, Campus Universitario, E-28871 Alcal\'a de Henares, Madrid, Spain}
\affil{$^{15}$School of Physics and Astronomy, Cardiff University, The Parade, Cardiff CF24 3AA, UK}
\affil{$^{16}$Istituto Nazionale di Astrofisica - Osservatorio Astronomico di Bologna, via Ranzani 1, I-40127 Bologna, Italy}
\affil{$^{17}$ESO, Karl-Schwarzschild-Str. 2, D-85748 Garching bei M\"unchen, Germany}
\affil{$^{18}$Subaru Telescope, 650 North A'ohoku Place, Hilo, HI 96720, USA}
\affil{$^{19}$National Astronomical Observatory of Japan, 2-21-1 Osawa, Mitaka, Tokyo 181-8588, Japan}
\affil{$^{20}$Department of Astronomy, School of Science, SOKENDAI (The Graduate University for Advanced Studies), 2-21-1 Osawa, Mitaka, Tokyo 181-8588, Japan}
\affil{$^{21}$Institute of Astronomy, The University of Tokyo, 2-21-1 Osawa, Mitaka, Tokyo 181-0015, Japan}
\affil{$^{22}$Graduate School of Science, The University of Tokyo, 7-3-1 Hongo, Bunkyo-ku, Tokyo 113-0033, Japan}
\affil{$^{23}$Dipartimento di Fisica e Astronomia, Universit\'a di Bologna, viale Berti Pichat 6/2, 40127 Bologna, Italy}
\affil{$^{24}$Department of Physics \& Astronomy, University of British Columbia, 6224 Agricultural Road, Vancouver, BC V6T 1Z1, Canada}
\affil{$^{25}$Ritter Astrophysical Research Center, University of Toledo, Toledo, OH 43606, USA}
\affil{$^{26}$Kapteyn Astronomical Institute, University of Groningen, Postbus 800, 9700 AV, Groningen, The Netherlands }
\affil{$^{27}$SRON Netherlands Institute for Space Research, Landleven 12, 9747 AD, Groningen, The Netherlands }
\affil{$^{28}$Department of Physics and Astronomy, University of the Western Cape, Robert Sobukwe Road, 7535 Bellville, Cape Town, South Africa }
\affil{$^{29}$INAF -- Istituto di Radioastronomia, Via Gobetti 101, 40129 Bologna, Italy}
\affil{$^{30}$Dipartimento di Fisica e Astronomia, Alma Mater Studiorum, Universit\`a degli Studi di Bologna, Via Gobetti 93/2, 40129 Bologna, Italy}
\affil{$^{31}$INAF -- Osservatorio Astronomico di Bologna, Via Gobetti 93/3, 40129 Bologna, Italy}
}
\jid{PASA}
\doi{10.1017/pas.\the\year.xxx}
\jyear{\the\year}

\usepackage[authoryear]{natbib}
\bibpunct{(}{)}{;}{a}{}{,}
\usepackage{aas_macros}
\usepackage{hyperref} 
\hypersetup{colorlinks,citecolor=blue,linkcolor=blue,urlcolor=blue}

\begin{document}%
\begin{abstract}
The mid-infrared (IR) range contains many spectral features associated with large molecules and dust grains such as polycyclic aromatic hydrocarbons (PAHs) and silicates. These are usually very strong compared to fine-structure gas lines, and thus valuable in studying the spectral properties of faint distant galaxies. In this paper, we evaluate the capability of low-resolution mid-IR spectroscopic surveys of galaxies that could be performed by SPICA. The surveys are designed to address the question how star formation and black hole accretion activities evolved over cosmic time through spectral diagnostics of the physical conditions of the interstellar/circumnuclear media in galaxies. On the basis of results obtained with {\it Herschel} far-IR photometric surveys of distant galaxies and {\it Spitzer} and AKARI near- to mid-IR spectroscopic observations of nearby galaxies, we estimate the numbers of the galaxies at redshift $z > 0.5$, which are expected to be detected in the PAH features or dust continuum by a wide (10 deg$^2$) or deep (1 deg$^2$) blind survey, both for a given observation time of 600 hours. As by-products of the wide blind survey, we also expect to detect debris disks, through the mid-IR excess above the photospheric emission of nearby main-sequence stars, and we estimate their number. We demonstrate that the SPICA mid-IR surveys will efficiently provide us with unprecedentedly large spectral samples, which can be studied further in the far-IR with SPICA.  
\end{abstract}
\begin{keywords}
galaxies: evolution -- galaxies: star-formation -- galaxies: active -- infrared: galaxies -- infrared: ISM -- methods: observational
\end{keywords}
\maketitle%

{\bf Preface}

\vspace{0.5cm}
\noindent
The following set of papers describe in detail the science goals of the future Space Infrared telescope for Cosmology and Astrophysics (SPICA). The SPICA satellite will employ a 2.5-m telescope, actively cooled to around 6\,K, and a suite of mid- to far-IR spectrometers and photometric cameras, equipped with state of the art detectors. In particular the SPICA Far Infrared Instrument (SAFARI) will be a grating spectrograph with low ($R = 300$) and medium ($R \simeq 3000$--11000) resolution observing modes instantaneously covering the 35--230\,$\mu$m wavelength range. The SPICA Mid-Infrared Instrument (SMI) will have three operating modes:  a large field of view ($12'\times10'$) low-resolution 17--36\,$\mu$m spectroscopic ($R \sim 50$--120) and photometric camera at 34\,$\mu$m, a medium resolution ($R \simeq 2000$) grating spectrometer covering wavelengths of 18--36\,$\mu$m and a high-resolution echelle module ($R \simeq 28000$) for the 12--18\,$\mu$m domain.  A  large field of view ($80''\times80''$), three channel, (110\,$\mu$m, 220\,$\mu$m and 350\,$\mu$m) polarimetric camera will also be part of the instrument complement. These articles will focus on some of the major scientific questions that the SPICA mission aims to address, more details about the mission and instruments can be found in \citet{roe17}.

\section{Introduction }
One of the biggest questions in current astrophysical research is how star formation and black hole accretion activities evolved throughout cosmic history. In order to answer the question, we need efficient methods to study the spectral properties of a large sample of galaxies in a systematic way, and thereby trace not only those activities over cosmic time but also the profound relationship between the two phenomena through spectral diagnostics of the physical conditions of the interstellar/circumnuclear media in galaxies. It is particularly important to cover the peak phases of the two phenomena, which occur in the redshift range of $z = 1$--$3$ \citep{madau14}, and that redshift range corresponds to the cosmic time where dust extinction is most severe, making any UV, optical and near-infrared (IR) observations prone to large systematic errors which render the results highly unreliable. X-rays are useful for detecting AGN, but can miss the population of Compton-thick AGN. The mid- to far-IR spectral range contains an enormous number of ionic, atomic and molecular lines and dust features as spectral dianostic tools \citep[e.g.,][]{spinoglio92}. Hence IR spectroscopic surveys from space are crucial. 

More specifically, in the mid-IR range, there are many important spectral bands of dust particles and very large molecules such as silicates, carbonaceous grains, and ices. Among them, emission features due to polycyclic aromatic hydrocarbons (PAHs) are ubiquitously observed from photo-dissociation regions (PDRs), which are widely distributed around star-forming regions in a galaxy \citep[e.g.,][]{hollenbach99}. The PAH emission is also detected from the diffuse interstellar medium. Their emission features are detected not only from many nearby galaxies \citep[e.g.,][]{smith07} but also from distant galaxies up to redshift $z\sim 4$ \citep{yan07, sajina12, riechers14, kirkpatrick15}. PAHs are believed to be the most important heating agents of gas in PDRs \citep[e.g.,][]{weingartner01}, and thus their emissions are crucial probes to study the interstellar media associated with star-formation activity. In particular, PAH spectral features at 3.3, 6.2, 7.7, 8.6, 11.3, 12.7 and 17~$\mu$m, which are attributed to C-C stretching and C-H bending modes, are notably strong compared to fine-structure gas lines for star-forming galaxies, though PAH spectral features are relatively broad. Hence they are powerful tools to determine the redshifts of faint distant galaxies, and can also trace star-formation activity since PAH features are characteristic of PDRs \citep[e.g.,][]{lutz08, teplitz07, takagi10, bonato15, shipley16}.

The PAH features are also useful to estimate the relative
contribution of an active galactic nucleus (AGN) and the star
formation component to the total IR luminosity $L_{\rm IR}$ of a
galaxy; the emission from PAHs is expected to be suppressed 
by photo-dissociation of PAHs due to the hard UV and X-ray radiation
field from the AGN, while the mid-IR continuum emission is enhanced 
because of heating of circumnuclear dust by the same radiation
\citep{oyabu11,lacy13}. Therefore the equivalent widths of the PAH
features enable us to roughly estimate the star-formation contribution
to the total $L_{\rm IR}$ of a galaxy \citep[][]{moorwood86,roche91,genzel98,armus07,imanishi07,imanishi08,imanishi10,veilleux09, nardini08,nardini09,nardini10, pope08, menendez09, coppin10, stierwalt13, stierwalt14}, 
and more reliably estimate it when they are normalized
with other spectral indicators (such as H$_2$ or [NeII] 12.8~$\mu$m line fluxes) or the slope of the IR continuum \citep{tommasin10}. This approach is complementary to the one that uses spectral energy distribution (SED) fitting \citep[e.g.,][]{gruppioni16, delvecchio14}.
%
%
%{\bf The PAHs are bright readily identified, their band positions are stable, 
%and the variation of the interband ratios are only factors of $\sim$2 \citep{smith07} ,}
%
%The PAH interband ratios are known to be similar from galaxy to galaxy \citep{smith07},
%and that stability is particularly useful in estimating star-formation rates. 
Although the PAH features are bright and readily identified, it is also
known that their interband ratios can vary from galaxy to galaxy to some
extent, mainly depending on their ionization states and/or size
distributions which reflect the interstellar conditions
\citep[e.g.,][]{allamandola89,joblin94}. Typical examples in the nearby
universe are PAHs in early-type galaxies, where the PAH 6.2 and 7.7~$\mu$m features are significantly weaker than the PAH 11.3~$\mu$m
feature \citep{kaneda05,kaneda08,panuzzo11}. The profiles of the PAH
features, such as peak positions, widths and relative strengths of
plateau components, might also vary \citep[e.g.,][]{tielens08},
providing us with information on the properties of the interstellar
medium in a galaxy (e.g., aromatic/aliphatic ratios). In addition to the variations of the PAH features, complications from metallicity may be a serious issue, especially when we discuss galaxies at high redshift; the abundance of PAHs relative to dust is known to decrease significantly at low metallicities \citep{engelbracht08}.
 
The strong silicate features at 9.7 and 18~$\mu$m are often detected
from a galaxy as either absorption or emission features. Similarly to
the PAH features, the silicate bands provide us with information not
only on the amount of silicate dust, but also on its properties such as
size distributions, crystallinity, and the degree of processing
\citep[e.g., porosity, Fe/Mg, olivine/pyroxene;][]{henning10,xie17}. In
particular the silicate features are the cornerstone of the AGN torus
paradigm. Their profiles range from moderate emission, usually but not
exclusively, in type 1 AGN, to deep absorption in the most dust-obscured
AGN. The strength of the silicate feature is indicative of the optical
depth of the hot dust heated by the active nucleus, and the relative
strength between the features at 9.7 and 18~$\mu$m provides information
on the distribution of the dust, as a smooth or clumpy medium
\citep{hatzim15}. Furthermore the combined information provided by the
PAH and silicate features allows for an almost unbiased classification
of objects into starburst- and AGN-dominated in the mid-IR
\citep{spoon07,hernan11}. Above all, both PAH and silicate features in
the mid-IR are not mere tools to estimate star-formation rates and 
AGN contribution in a galaxy, but also important probes
to study the physics governing the interstellar/circumnuclear media in a
galaxy. The potential of those dust features as spectral diagnostics,
however, is still not completely developed even in the nearby universe, much less at high redshift.

SPICA (SPace Infrared Telescope for Cosmology and Astrophysics), a 2.5-m
large cryogenic telescope in space, will provide unprecedented high
spectroscopic sensitivities with continuous wavelength coverage from the
mid- to the far-IR \citep[Roelfsema in prep.;][]{nakagawa14}. In particular
an extremely low IR background achieved thanks to its primary
mirror cooled down to $\sim 8$~K, is essential to study broad spectral
features such as dust bands from faint objects. SMI \citep[SPICA Mid-infrared Instrument;][]{kaneda16} is one of the two focal-plane scientific instruments planned for SPICA. 
SMI is the Japanese-led instrument proposed and managed by a university
consortium, designed to provide a longer wavelength coverage and higher spectral mapping efficiency (i.e., higher spectral survey speed) compared to JWST (James Webb Space Telescope), in addition to high-resolution spectroscopic capability. In this paper, we focus on the scientific potential of unbiased large spectroscopic surveys with SMI. On the other hand, \citet{gruppioni17} highlight the potential of large photometric surveys with SMI; they also describe the general scientific values of SPICA mid-IR survey datasets to reveal the evolution of the dust-obscured star-formation and AGN activity in galaxies since the re-ionization epoch at $z\sim 7$. We plan to perform follow-up spectroscopy with the SPICA far-IR instrument, SAFARI, based on the results of the SMI surveys, which is essential to complete our IR spectroscopic studies of the evolutions of galaxies and materials therein.

Throughout this paper, we adopt the flat universe with the following cosmological parameters: Hubble constant, $H_0=70\ \mathrm{km}\ \mathrm{s}^{-1}\ \mathrm{Mpc}^{-1}$, density parameter, $\Omega_{\rm M}=0.3$,  and cosmological constant, $\Omega_{\rm \Lambda}=0.7$.

\section{SPICA Mid-infrared Instrument (SMI) for large surveys}
\begin{figure}[htbp]
\begin{center}
\includegraphics[height=\columnwidth,angle=270]{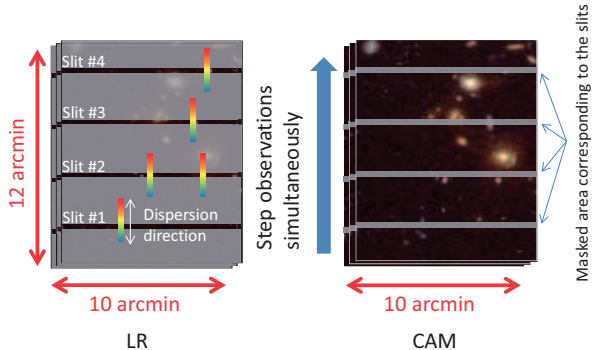}
\caption{Schematic images of the SMI/LR multi-slit spectroscopic survey with the SMI/CAM slit viewer for pointing reconstruction in creating low-resolution spectral maps. A spatial scan with 90 steps produces a spectral map of $10' \times 12'$ area.}
\label{fig1}
\end{center}
\end{figure}

SMI has the following four channels in the mid-IR: spectroscopic
functions for low-resolution (LR), mid-resolution (MR) and
high-resolution (HR) spectroscopy, and a photometric function for
broad-band imaging (CAM). The main design driver for SMI/LR and /CAM 
is the ability to carry out large surveys, especially of PAH spectra 
\citep{wada17}; a pioneering PAH spectral survey was performed 
by \citet{bertincourt09} with {\it Spitzer}/IRS.
SMI/LR is a multi-slit prism spectrometer system
with a wide field-of-view covered by 4 long slits of $10'$
in length and $3.7''$ in width, thus enabling low-resolution ($R$ =
50--120) spectroscopic surveys with continuous coverage of the
wavelength range of $17$--$36$~$\mu$m. In the SMI/LR system, a $10' \times
12'$ slit viewer camera (SMI/CAM) is implemented to accurately determine
the positions of the slits on the sky for pointing reconstruction in
creating spectral maps. SMI/CAM adopts an optical bandpass ($30$--$37$~$\mu$m) filter at a central wavelength of 34~$\mu$m and thus provides 34~$\mu$m broad-band images with a field of view of $10' \times 12'$ excluding the positions of the 4 slits. The design of the slit viewer in SMI/LR is based on the step-scan mode strategy implemented for large surveys. SMI/LR would produce a spectral map of $10' \times 12'$ area as a minimum field unit for a spatial scan with 90 steps (1 step length $\sim 2''$, i.e., half a slit width). Figure \ref{fig1} explains the concept of the SMI/LR spectral mapping method; the multi-slit spectrometer LR and the slit viewer CAM are operated simultaneously, providing multi-object low-resolution spectra at $17$--$36$~$\mu$m and broad-band deep images at 34~$\mu$m, respectively. 

\section{Survey strategy}
\begin{figure}[htbp]
\begin{center}
\includegraphics[width=\columnwidth]{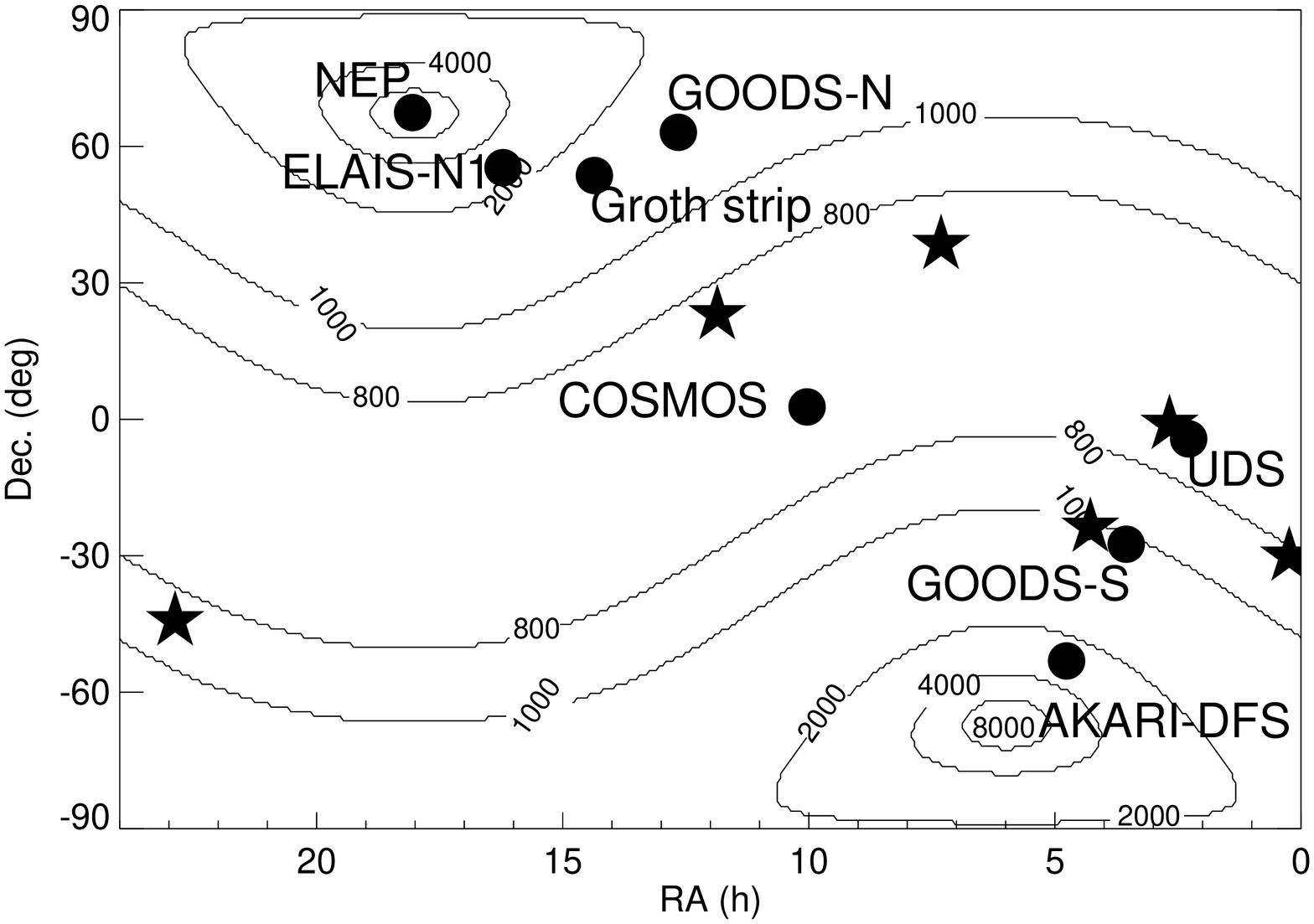}
\caption{Sky visibility contours of SPICA in units of hours per year. Circles identify popular extragalactic survey fields, NEP \citep[North Ecliptic Pole;][]{nep1,nep2}, ELAIS-N1 \citep[European Large Area ISO Survey;][]{elais}, Groth strip \citep{groth}, GOODS-N/S \citep[Great Observatories Origins Deep Survey;][]{goods}, COSMOS \citep[Cosmic Evolution Survey;][]{cosmos}, UDS \citep[Ultra Deep Survey;][]{uds}, and AKARI-DFS\citep[AKARI Deep Field South;][]{dfs,baronchelli16}. Stars indicate the Hubble Space Telescope Frontier Fields \citep{lotz17}, which are known to contain high-magnification gravitational lensing clusters of galaxies. We assume that SPICA can observe a $\pm 8$ degree zone along a great circle perpendicular to the solar vector.}
\label{fig2}
\end{center}
\end{figure}

As reference surveys, we consider two blind spectroscopic
surveys with SMI/LR: i) a wide survey of a 10 deg$^{2}$ area aimed
at covering various galaxy environments across the cosmic large-scale
structure, including (proto-)clusters of galaxies, as well as
serendipitously detecting rare luminous galaxies; and ii) a deep survey of a 1 deg$^{2}$ area that will cover a wide range of $L_{\rm IR}$ down to those of ordinary star-forming galaxies \citep[i.e., star-formation main-sequence galaxies;][]{elbaz11} at redshift $z\sim 3$. Table \ref{tab:survey} summarizes the parameters of these reference surveys, where the total observation time is assumed to be 600 hours in both cases. 600 hours is chosen to fit the time allocation plan based on the reference mission scenario for SPICA %\citep{roelfsema17}.
(Roelfsema et al. in prep.).
The two surveys can be combined by any appropriate ratio while the total time is kept to be the same.
For each pointing of the spatial scan, we take into account a 20~s stabilization time of SPICA as an overhead.
To determine the on-source time, we multiplied the exposure time per
step by a factor of 1.5, considering the overlap between each
field-of-view by half a slit width. Most of the popular
extragalactic survey fields have enough visibility for the assumed
observational time of 600 hours (see the sky visibility for SPICA in Figure \ref{fig2}). Hence the whole areas of 10 deg$^{2}$ and 1 deg$^{2}$ can be covered either contiguously or separately in principle, but a contiguous mapping would be one of our key advantages over JWST to enhance the ability to perform clustering analyses. 

The estimation of the survey spectral sensitivity is based on the latest specifications of SMI/LR, e.g., 5$\sigma$; 1-hour continuum sensitivities of 25 and 60~$\mu$Jy at 20 and 30~$\mu$m, respectively \citep{sakon16}. SMI/CAM has a 5$\sigma$; 1-hour sensitivity of 13~$\mu$Jy; the imaging data obtained simultaneously for the wide and deep surveys have the detection limits of 11 and 3~$\mu$Jy, respectively. 
Scientifically, SMI/LR spectral data are particularly useful to study star-forming galaxies with the PAH features, while SMI/CAM imaging data are useful to probe dusty AGN by combining other wavelength data as well as the SMI/LR data themselves. As shown below, the SMI/LR surveys will provide so many ($\sim 10^5$) PAH spectra of galaxies that we can statistically examine PAH band variations as spectral diagnostics. Technically, given that the absolute flux in the 34~$\mu$m band is well calibrated, the spectral data with SMI/LR can be calibrated relative to SMI/CAM at 34~$\mu$m as an anchoring point.

The detection limits of the SMI/CAM imaging data are comparable to or less than the confusion limit for SPICA's 2.5-m diameter telescope \citep[9~$\mu$Jy at 34~$\mu$m;][]{gruppioni17}. It will be possible to recover fluxes even three times lower than the confusion limit by taking advantage of ancillary data at other wavelengths (preparatory and/or follow-up observations) that will allow us to precisely constrain the position. On the other hand, the SMI/LR spectral data have continuum detection limits of 380 and 110~$\mu$Jy at 30~$\mu$m for the wide and deep surveys, respectively, and thus the confusion makes only very small ($<20$\%) contributions to the underlying continuum in estimating the equivalent widths of dust features. As shown below, in the case of the SMI/LR deep survey, the population density of the detected galaxies reaches $2 \times 10^{-2}$ per beam ($3.7''$), which indicates that 2\% of the galaxies may be blended with another detected galaxy under the assumption that galaxies are uniformly distributed. From the data affected by the blending, it will be possible to extract the PAH spectrum of each galaxy by using the difference in their redshifts, but it will be difficult to recover their continuum components.

It should be noted that, in terms of the limiting flux density for SMI/CAM, the wide and deep surveys are almost equivalent to, and thus consistent with, the Deep Survey (DS) and the Ultra-Deep Survey (UDS) described in \citet{gruppioni17}, respectively. For the 100 deg$^2$ Shallow Survey (SS) proposed in \citet{gruppioni17}, we plan to conduct additional dedicated photometric surveys using only SMI/CAM, since the spatial step scan for SMI/LR would require an exposure time of $\sim 0.2$ s which is too short for the detectors to be operated.  

\begin{table}
\caption{Survey parameters}
\begin{center}
\begin{tabular}{lrr}
\hline\hline
Parameters & Wide & Deep \\
\hline
Survey area (deg$^2$)  & 10  & 1 \\
Number of fields ($10'\times12'$ as a unit) & 300  & 30 \\
Time per field w/o overheads (hours) & 1.45 & 18.85 \\
Total time incl. overheads (hours) & 600  & 600 \\
On-source time (sec) & 90  & 1170 \\
\hline\hline
\end{tabular}
\end{center}
\label{tab:survey}
\end{table}

\section{Measurement of PAH and dust emission fluxes}
\begin{figure}[htbp]
\begin{center}
\includegraphics[width=\columnwidth]{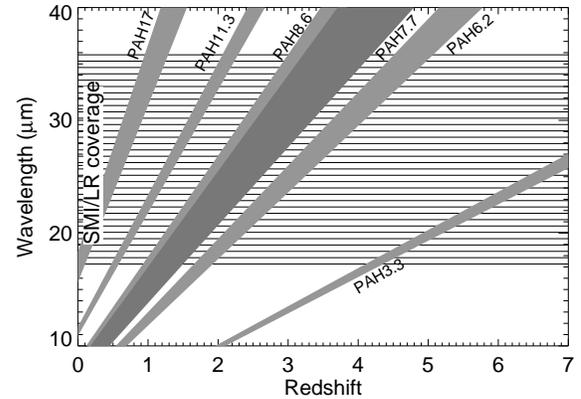}
\caption{Observed wavelengths of the PAH features as a function of redshift. The striped area indicates the wavelength range covered by SMI/LR.}
\label{fig:z-pah}
\end{center}
\end{figure}

We first estimate the limiting fluxes of the PAH features, based on the
current specifications of SMI/LR \citep{kaneda16,sakon16}. In the rest
frame, we adopt the approximate band widths ($\Delta\lambda$) of 0.08,
0.3, 0.9, 0.4, and 1.4~$\mu$m for the PAH features at 3.3, 6.2, 7.7,
11.3, and 17~$\mu$m, respectively \citep{draine07}. In the observed
frame, the band widths as well as the central wavelengths of the PAH
features change with redshift (Figure~\ref{fig:z-pah}); from the SMI/LR continuum sensitivities, their limiting fluxes in one hour (5$\sigma$) are calculated to be $8.2 \times 10^{-17}$, $1.2 \times 10^{-16}$, $1.8 \times 10^{-16}$, $1.0 \times 10^{-16}$, and $1.5 \times 10^{-16}$~erg/s/cm$^2$ at 20~$\mu$m, while they are $1.3 \times 10^{-16}$, $1.9 \times 10^{-16}$, $2.9 \times 10^{-16}$, $1.6 \times 10^{-16}$, and $2.4 \times 10^{-16}$~erg/s/cm$^2$ at 30~$\mu$m. Thus the limiting PAH fluxes vary with redshift by a factor of 1.6 within a range of the redshift where the central wavelength of the corresponding PAH feature is observed at 20~$\mu$m or 30~$\mu$m. For simplicity, we adopt the average of the above two fluxes as the one-hour limiting flux $f_{\rm limit}$ for each PAH feature in the following calculation (i.e., $f_{\rm limit} = 1.1 \times 10^{-16}$, 1.5 $\times 10^{-16}$,  2.4 $\times 10^{-16}$, 1.3 $\times 10^{-16}$, and $2.0 \times 10^{-16}$~erg/s/cm$^2$ for the PAH 3.3, 6.2, 7.7, 11.3, and 17~$\mu$m features, respectively).

For pure AGN, we assume that the PAH emission is faint and
non-detectable \citep{moorwood86, roche91}; in order to estimate
the numbers of the AGN expected to be detected in the SMI spectral surveys, we may be able to utilize the silicate emission
features \citep{hao05, sturm05}. However, unlike PAH, the silicate
features can be in either emission or absorption (or both), and also
their intrinsic strengths relative to $L_{\rm IR}$ are known to be
highly variable from object to object \citep{hatzim15,xie17}. Instead we
use 6~$\mu$m continuum emission of hot dust which is typical of AGN
just for the purpose of quantitative estimation; the silicate features
themselves would be scientifically important to further characterize
AGN. To convert the rest-frame 6~$\mu$m AGN continuum to that at the
observed wavelength, we assume that $\nu F_{\nu}$ is constant. This
assumption is reasonable for type 1 AGN because their dust continua are
relatively flat around 10--30~$\mu$m, exhibiting a broad peak in this range \citep{shang11}, and still acceptable for type 2 AGN because of similarity in continuum shapes at $\sim 3$--$30$~$\mu$m between type 1 and 2 quasars except for the silicate and PAH features \citep{hiner09}. Similar to the limiting PAH fluxes, from the SMI/LR and /CAM continuum sensitivities, the one-hour limiting fluxes (5$\sigma$) for hot dust emission ($\nu F_{\nu}$) are 3.7 $\times 10^{-15}$~erg/s/cm$^2$ at 20 $\mu$m and 6.0 $\times 10^{-15}$ erg/s/cm$^2$ at 30~$\mu$m for SMI/LR, while they are 1.1 $\times 10^{-15}$ erg/s/cm$^2$ for SMI/CAM at 34~$\mu$m. For SMI/LR, we again adopt the averages of the two fluxes as the one-hour limiting flux $f_{\rm limit}$ for AGN hot dust emission.

Since pure star-forming galaxies or pure AGN are somewhat extreme cases, we also consider a mixture of them. In the following calculation, we define the three types of galaxies, SF(Star Formation)100\%, AGN100\% galaxies, and SF50\%+AGN50\% galaxies. The last type corresponds to a galaxy where a half of $L_{\rm IR}$ is powered by star-formation activity while the other half is attributed to AGN activity. We consider their detections on the basis of the PAH features or the hot dust continuum emission, when we refer to them as star-forming galaxies (or PAH galaxies as defined below) or AGN, respectively.

For the wide and deep surveys with SMI/LR and /CAM, $f_{\rm limit}$ is scaled with the square root of the on-source exposure time (Table \ref{tab:survey}), since the SMI/LR sensitivity is limited by background photon noise. To convert $f_{\rm limit}$ to the limiting IR luminosity of a galaxy, $L_{\rm IR, limit}$, we use the following equation:
\begin{equation}
L_{\rm IR, limit} = 4\pi D_{\rm L}(z)^2\left(\frac{L_{\rm IR}}{L_{\rm PAH~or~hot~dust}}\right)f_{\rm limit}, 
\end{equation}
where $D_{\rm L}(z)$ is the luminosity distance. To simplify the
calculation, we assume that the luminosity of each of the PAH features
($L_{\rm PAH}$) and the hot dust continuum emission ($L_{\rm hot~dust}$)
is proportional to $L_{\rm IR}$. (In reality, their relative strengths,
especially PAH interband ratios, are expected to vary depending on the
properties of the interstellar medium in a galaxy, which is also to be
studied by SPICA.) The adopted PAH and the monochromatic continuum strengths at 6~$\mu$m relative to $L_{\rm IR}$ (i.e., proportionality coefficients) are summarized in Table \ref{tab:strength}, which are estimated from {\it Spitzer} and AKARI near- to mid-IR spectroscopic observations of nearby galaxies \citep{smith07,yamada13,nardini09}. For the intensities of the PAH 6.2, 7.7, 11.3, and 17~$\mu$m features in \citet{smith07}, we utilized the spline-fitting result but not the result of spectral decomposition (i.e., we excluded the contribution of PAH plateau components), consistently with the above assumption on the widths of the PAH features. For the 6~$\mu$m continuum strength, we adopted the value averaged for AGN in nearby ultra-luminous IR galaxies with various geometries of AGN tori \citep{nardini09}.
For SF50\%+AGN50\% galaxies, the relative strengths of the PAH features and the hot dust continuum are decreased by a factor of 2 from those of SF100\% and AGN100\% galaxies, respectively, as shown in Table \ref{tab:strength}. 

\begin{figure}[htbp]
\begin{center}
\includegraphics[width=\columnwidth]{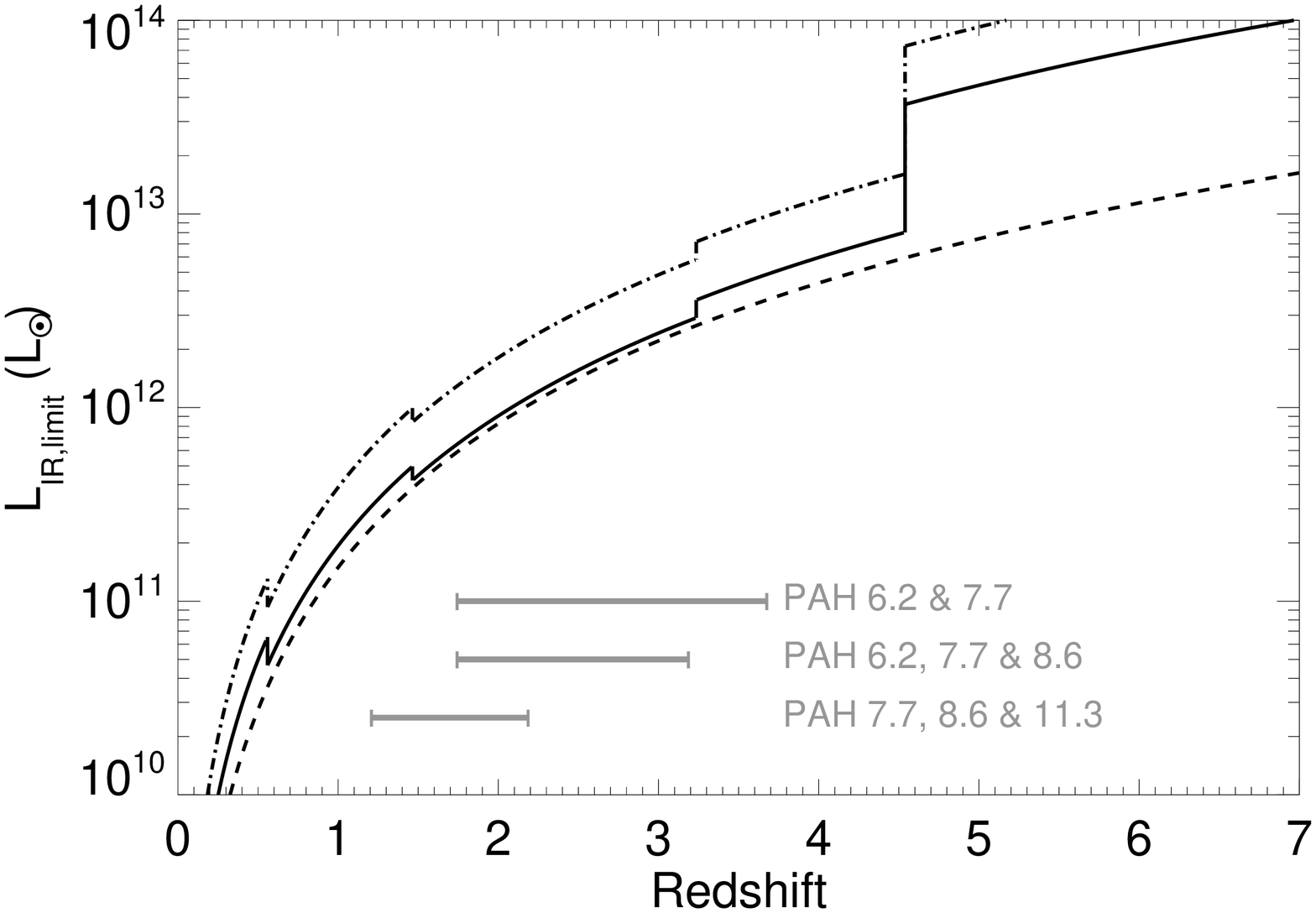}
\includegraphics[width=\columnwidth]{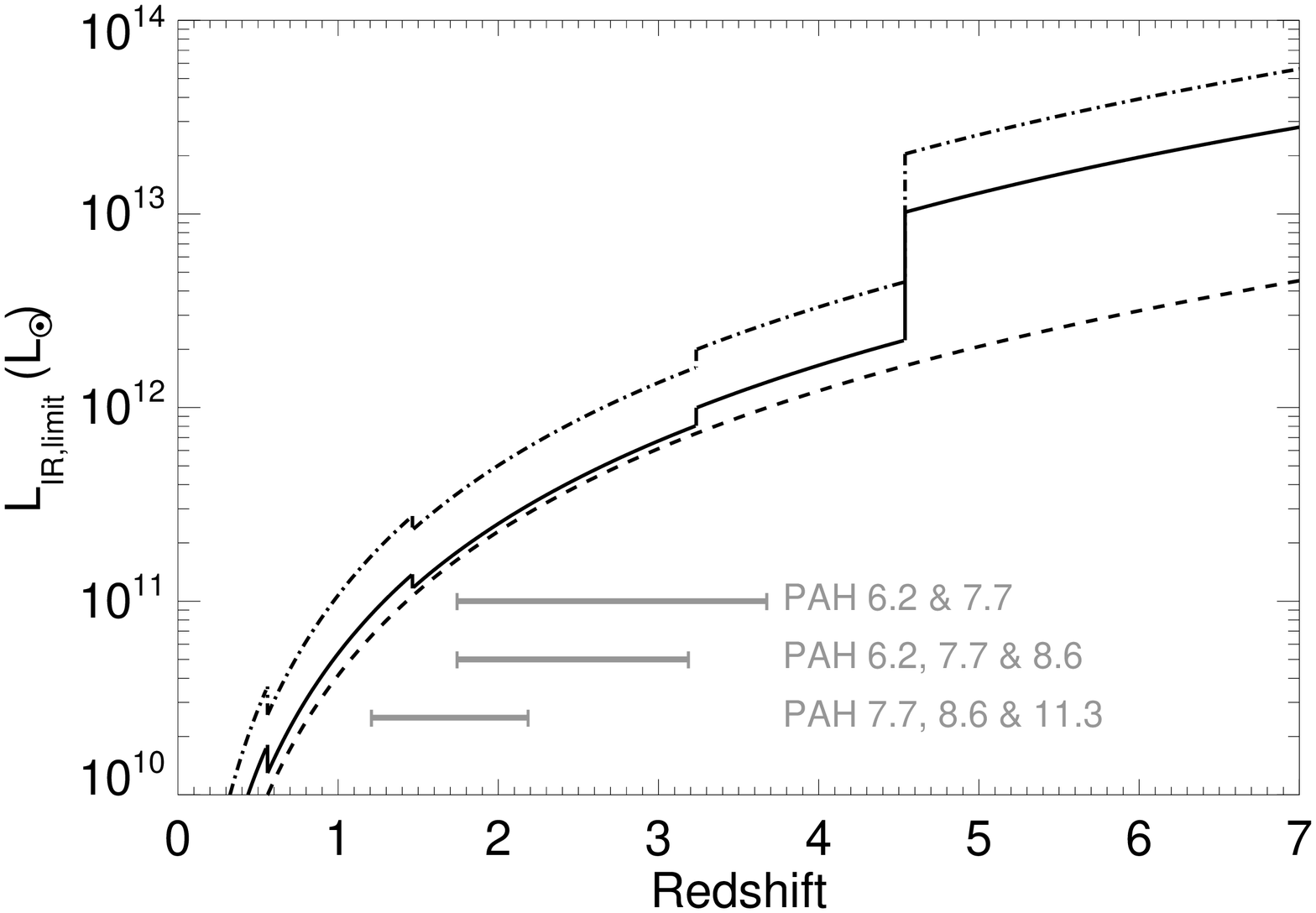}
\caption{Limiting IR luminosities $L_{\rm IR, limit}$ in Equation (1) for (top) the wide and (bottom) deep surveys, calculated as a function of redshift for the galaxies of SF100\% (solid line), SF50\%+AGN50\% (dot-dashed line) and AGN100\% (dashed line), while the horizontal bars show the redshift ranges where multiple major PAH features are available (i.e., their peaks are included).}
\label{fig3}
\end{center}
\end{figure}

The limiting IR luminosities, $L_{\rm IR, limit}$, are calculated as a function of redshift, $z$, for SF100\%, SF50\%+AGN50\%, and AGN100\% galaxies. Figure \ref{fig3} shows the resultant $L_{\rm IR, limit}$ for the wide and deep surveys, where $L_{\rm IR, limit}$ for SF50\%+AGN50\% galaxies is based on the PAH features. The discontinuity in the plots is caused by appearance or disappearance of the corresponding PAH feature in the SMI/LR spectral range. Here we consider an edge margin of $\pm\Delta\lambda$ for inclusion of each PAH feature in the SMI/LR range of $17$--$36$ $\mu$m. In the figure, from low to high redshift, the PAH 17, 11.3, 7.7, 6.2 and 3.3~$\mu$m features determine the behavior of $L_{\rm IR, limit}$ as a function of $z$. 

\begin{table}
\caption{Strengths of the PAH feature and 6~$\mu$m continuum luminosity relative to the total IR luminosity for SF100\%, AGN100\%, and SF50\%+AGN50\% galaxies}
\label{tab:strength}
\begin{center}
\begin{tabular}{crrr}
\hline\hline
  & \multicolumn{3}{c}{$\log \left( L_{\rm PAH~or~hot~dust}/L_{\rm IR} \right)$}\\
  & SF100\% & AGN100\% & SF50\%+\\
  &         &          & AGN50\%\\
\hline
PAH 3.3~$\mu$m $^1$            &  $-$3.0 & --- &  $-$3.3 \\
PAH 6.2~$\mu$m $^2$            &  $-$2.2 & --- &  $-$2.5 \\
PAH 7.7~$\mu$m $^2$            &  $-$1.9 & --- & $-$2.2 \\
PAH 11.3~$\mu$m $^2$           &  $-$2.2 & --- & $-$2.5 \\
PAH 17~$\mu$m $^2$             &  $-$2.2 & --- & $-$2.5 \\
$\nu L_{\nu} $ at 6~$\mu$m $^3$ &  $-$1.9 & $-$0.55 & $-$0.83\\
\hline
\end{tabular}
\end{center}
\tabnote{$^1$ \citet{yamada13}}
\tabnote{$^2$ Spline-fitting result in \citet{smith07}}
\tabnote{$^3$ \citet{nardini09}}
\end{table}

\section{Expected results}
\subsection{Numbers of galaxies}
\begin{figure}[htbp]
\begin{center}
\includegraphics[width=\columnwidth]{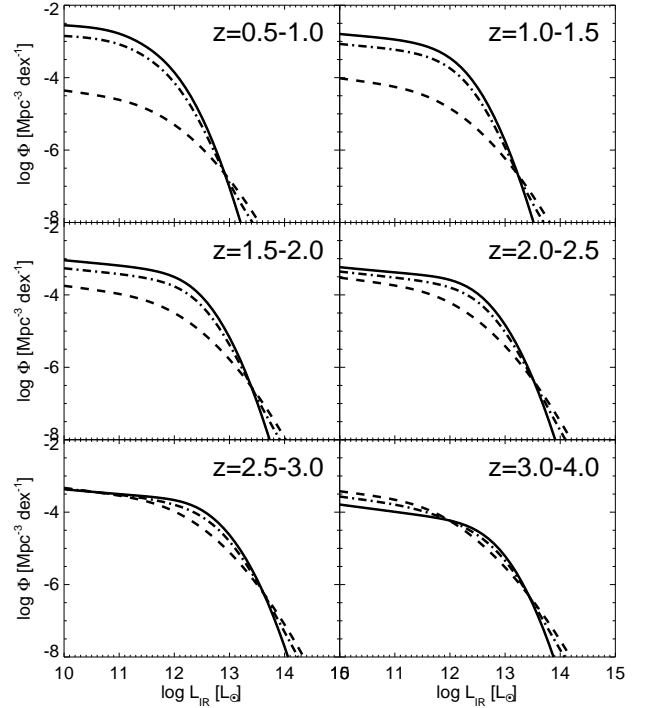}
\caption{Luminosity functions for the galaxies of SF100\%(solid line),
  SF50\%+AGN50\%(dot-dashed line) and AGN100\%(dashed line), calculated for the parameters given in \citet{gruppioni13}.}
\label{fig:lf}
\end{center}
\end{figure}

The luminosity functions used in our calculation are based on those
given in \citet{gruppioni13} with the {\it Herschel} far-IR surveys, where
they define the five populations of spiral, starburst, SF-AGN, AGN1 and
AGN2, and give the parameters of the luminosity functions for each
population. We define the above three types as SF100\% =
2/3$\times$(spiral + starburst + SF-AGN), AGN100\% = 2/3$\times$(AGN1 +
AGN2), and SF50\%+AGN50\% = 1/2$\times$(SF100\% + AGN100\%), so that the
total numbers are conserved. Note that, by definition, our SF50\%+AGN50\% is closer to AGN2, while their SF-AGN has the IR luminosity dominated by SF \citep{gruppioni13}. 
As a result, we assume significantly more SF-AGN composite systems (i.e., 33\% of the total) than suggested in \citet{gruppioni13} (e.g., $\sim$15\% at $z \sim 3$), which would give conservative values on the numbers of PAH galaxies in the calculation below. 
Figure \ref{fig:lf} shows the luminosity
functions thus derived for different redshift ranges. Above $z=4$, we
assume the same luminosity functions as those at $z = 3$--$4$, since the parameters of the luminosity functions in \citet{gruppioni13} are not well constrained in that redshift range. 

The number of galaxies at redshifts between $z_1$ and $z_2$ with luminosity between $L_1$ and $L_2$ is derived as follows:
\begin{equation}
N=\Omega\int_{z_1}^{z_2}\int_{\log L_1}^{\log L_2}\Phi(z,L_{\rm IR})\frac{dV(z)}{dzd\Omega}d\log L_{\rm IR}dz,
\end{equation}
where $\Omega$ is the solid angle of the survey area (i.e., $\Omega =
3.05\times 10^{-3}$ and $3.05\times 10^{-4}$ sr for the wide and deep
surveys, respectively), $\Phi(z,L_{\rm IR})$ is the luminosity function
at redshift $z$ and luminosity $L_{\rm IR}$, and $dV(z)/dzd\Omega$ is
the comoving volume in a redshift interval $dz$ within a solid angle $d\Omega$. The comoving volume is calculated by adopting the flat universe with the cosmological parameters listed above. If $L_1$ is fainter than $L_{\rm IR, limit}$ which is estimated in the previous section, we use $L_{\rm IR, limit}$ instead of $L_1$. 

\begin{figure}[htbp]
\begin{center}
\includegraphics[width=\columnwidth]{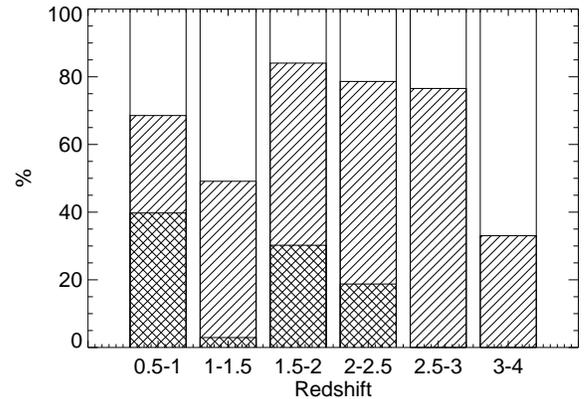}
\caption{Fractions of the galaxies detected with 2 (stripe) and 3 PAH features (cross) for the wide survey.}
\label{fig:fraction}
\end{center}
\end{figure}

Tables \ref{tab:pahwide} and \ref{tab:pahdeep} summarize the numbers of PAH galaxies expected to be detected per bin of the $L_{\rm IR}-z$ plane for the SMI/LR wide and deep surveys, respectively. Here we define the PAH galaxies as a sum of SF100\% and SF50\%+AGN50\% galaxies, and show the contribution of the SF50\%+AGN50\% galaxies in the parentheses. Considering that typical variations of the PAH feature strengths relative to $L_{\rm IR}$ are $\sim \pm 30$\% with 1$\sigma$ for star-forming galaxies of a near-solar metallicity and in the absence of AGN \citep{smith07}, we estimate the effect of such variations by changing $L_{\rm IR}/L_{\rm PAH}$ in Equation (1) by $\pm$30\%, and find that the total numbers of PAH galaxies in Tables \ref{tab:pahwide} and \ref{tab:pahdeep} can vary by $\sim 20$\% at lower redshift to $\sim 40$\% at higher redshift. Figure~\ref{fig:fraction} shows the fractions of the PAH galaxies detected with two or three PAH features among those at 6.2, 7.7, 8.6, 11.3, 12.7 and 17~$\mu$m. As can be seen in the figure, a majority of the PAH galaxies are detected with multiple PAH features for $z \le 3$, which is important in order to accurately determine the redshift (see Section 5.2).

\begin{figure}[htbp]
\begin{center}
\includegraphics[width=\columnwidth]{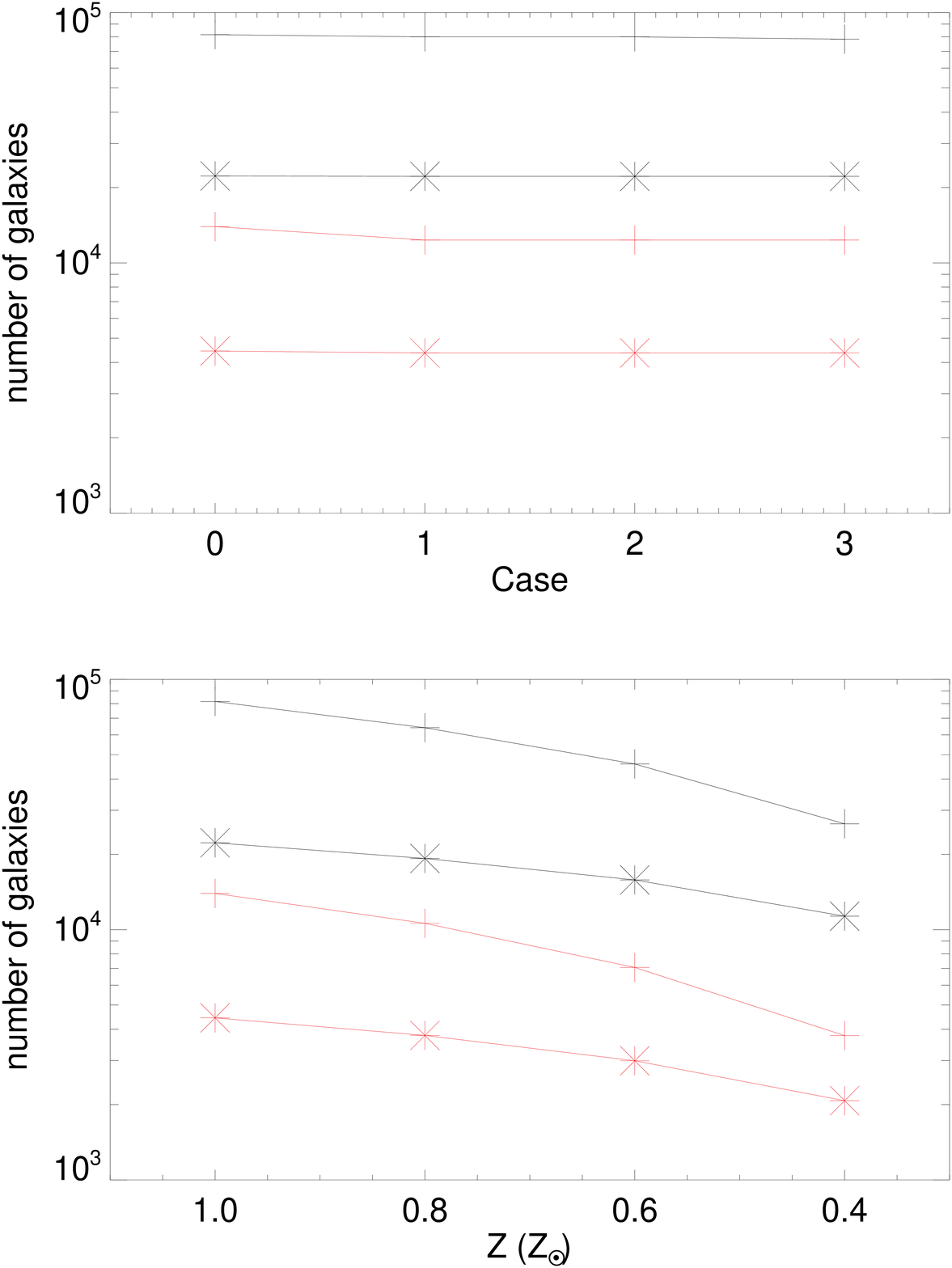}
\caption{Effects of systematic changes of $L_{\rm PAH} / L_{\rm IR}$ as a function of (top) $L_{\rm IR}$, redshift, and (bottom) metallicity on the total numbers of PAH galaxies for the wide (black pluses) and deep (black asterisks) surveys. Red symbols are those limited to $z>2$. (Top) Cases 1--3 correspond to different assumptions on $L_{\rm PAH} / L_{\rm IR}$ as a function of $L_{\rm IR}$ and redshift (see text for detail), while Case 0 is for the constant $L_{\rm PAH} / L_{\rm IR}$ assumption (i.e., Tables \ref{tab:pahwide} and \ref{tab:pahdeep}). (Bottom) The changes are calculated based on the metallicity dependence given in \citet{engelbracht08}.}
\label{fig:ng}
\end{center}
\end{figure}

In Figure~\ref{fig:ng}, we evaluate the effects of systematic changes of the PAH feature strengths relative to $L_{\rm IR}$. First we consider the fact that $L_{\rm PAH}/L_{\rm IR}$ decreases with $L_{\rm IR}$ for local ultra-luminous IR galaxies \citep[ULIRGs;][]{imanishi10,yamada13,desai07}, while high-$z$ ULIRGs tend to have higer $L_{\rm PAH}$ than local ULIRGs \citep[e.g.,][]{pope08,pope13}. Using the $L_{\rm PAH}/L_{\rm IR}$--$L_{\rm IR}$ relationships at $z = 0$ and $z = 2$ given in \citet{shipley16}, we assume the following three cases for the evolution with redshift, namely that the low-$z$ and high-$z$ relationships change at (1) $z = 0.5$, (2) 1.0 and (3) 1.5. Figure~\ref{fig:ng} (top) shows the results for the three cases, from which we find that the systematic changes of $L_{\rm PAH} / L_{\rm IR}$ as a function of $L_{\rm IR}$ and redshift do not significantly affect the numbers of PAH galaxies; most of the galaxies are not that IR-bright. Second we consider the fact that $L_{\rm PAH}/L_{\rm IR}$ decreases with metallicity \citep{engelbracht08}. Several works have reported that the metallicities of ULIRGs from the local universe to $z\sim$5 are $Z= 0.5$--$1.5$ Z$_{\odot}$ \citep{rupke08,pereira17,fadely10,wardlow17,nagao12,bethermin16}. LIRGs at $z \sim 2 $, which have the stellar mass of $M_{\mathrm{\odot}} > 10^{9.6}\ \mathrm{M}_{\odot}$ \citep{daddi07}, are expected to have the metallicity of $Z > 0.5\ \mathrm{Z}_{\odot}$ from the mass-metallicity relation at $z=2$ \citep{maiolino08}. Thus we assume three cases for metallicity, namely $Z= 0.8$, 0.6, and 0.4 Z$_{\odot}$. Figure~\ref{fig:ng} (bottom) shows the result for the effect of the metallicity in these cases, from which we find that the numbers of PAH galaxies can be reduced by a factor of 2--3. There are many other effects that can change $L_{\rm PAH}/L_{\rm IR}$; for example, it might be harder for PAHs to survive if there is less dust shielding and the radiation field is stronger at high $z$. Neverthess it should be noted that targets that are unique for SPICA are the dusty, obscured populations that are relatively rich with metals and dust.
 
%Considering that typical variations of the PAH feature strengths relative to $L_{\rm IR}$ are $\sim \pm 30$\% with 1$\sigma$ for star-forming galaxies of a near-solar metallicity and in the absence of AGN \citep{smith07}, we estimate the effect of such variations by changing $L_{\rm IR}/L_{\rm PAH}$ in Equation (1) by $\pm$30\%, and find that the total numbers of PAH galaxies in Tables \ref{tab:pahwide} and \ref{tab:pahdeep} can vary by $\sim 20$\% at lower redshift to $\sim 40$\% at higher redshift. 

On the other hand, Tables \ref{tab:agnwide} and \ref{tab:agndeep} summarize the numbers of AGN expected to be detected per bin of the $L_{\rm IR}-z$ plane for the SMI/LR wide and deep surveys, respectively, where we define the AGN as a sum of AGN100\% and SF50\%+AGN50\% galaxies. In the parentheses of the tables, we also show the numbers of the SF50\%+AGN50\% galaxies as AGN. Note that, to obtain the total numbers of galaxies, we add the values in the tables for the PAH galaxies and AGN, and then subtract the values in parentheses in the tables for the PAH galaxies. Finally Table \ref{tab:agnwidecam} lists the numbers of AGN expected to be detected with SMI/CAM in the surveys. 

The tables clearly show that a huge number of galaxies are
 foreseen to be detected in the SMI surveys. In particular, from Tables
\ref{tab:pahwide}, \ref{tab:agnwide} and \ref{tab:agnwidecam}, we find that the wide survey
would produce $\sim 5\times10^4$ spectra of PAH galaxies at $z>1$, among
which $\sim 1.4\times10^4$ spectra would come from galaxies at $z = 2$--$4$, as well as $\sim 2\times10^4$ spectra of AGN at $z>1$, while the slit viewer would detect more than $2\times10^5$ dusty AGN at $z>1$. On the other hand, Tables \ref{tab:pahdeep} and \ref{tab:agndeep} show that the deep survey reaches luminosity levels of so-called star-formation main sequence galaxies (e.g., $L_{\rm IR} \sim 1\times 10^{12}$~L$_{\odot}$ at $z\sim 3$) with a fair margin. Thus, with the sample size and depth in the tables, we will be able not only to establish robust rest-frame mid-IR spectral samples as a function of $z$ and $L_{\rm IR}$, but also to examine PAH and silicate band variations as spectral diagnostics of the physical conditions in the star-forming regions and the nuclei of galaxies with the help of other wavelength spectral data.

\begin{table*}[htbp]
  \begin{center}
  \caption{Numbers of the PAH galaxies (SF100\% and SF50\%+AGN50\%) expected to be detected in the SMI/LR wide survey. The values in the parentheses are the numbers of SF50\%+AGN50\% galaxies.}  \label{tab:pahwide}
\small
    \begin{tabular*}{\textwidth}{lrrrrrrr}
      \hline\hline
      &\multicolumn{6}{c}{Redshift}\\
      $\log(L_{\rm IR}/$L$_{\odot})$ & 0.5$-$1 & 1$-$1.5 & 1.5$-$2.0 & 2.0$-$2.5 & 2.5$-$3.0 & 3.0$-$4.0 & $>$ 4.0 \\
      \hline
13.00$-$   &     1 (    1) &    19 (    9)&    91 (   38)&   246 (  101)&   462 (  193)&   225 (   99)&   113 (   36)\\
12.50$-$13.00 &    73 (   27) &   648 (  231)&  1592 (  576)&  2495 (  934)&  2707 (  847)&   682 (   85)&    79 (    0)\\
12.25$-$12.50 &   321 (  112) &  1825 (  631)&  3009 ( 1068)&  2784 (  583)&  1659 (    0)&    67 (    0)&     0 (    0)\\
12.00$-$12.25 &  1138 (  390) &  4446 ( 1527)&  4535 (  974)&  2241 (    0)&   110 (    0)&     0 (    0)&     0 (    0)\\
11.75$-$12.00 &  2969 ( 1008) &  7057 ( 1677)&  3676 (   50)&    86 (    0)&     0 (    0)&     0 (    0)&     0 (    0)\\
11.50$-$11.75 &  5980 ( 1968) &  6496 (  438)&   406 (    0)&     0 (    0)&     0 (    0)&     0 (    0)&     0 (    0)\\
11.00$-$11.50 & 16502 ( 2877) &  2113 (    0)&     0 (    0)&     0 (    0)&     0 (    0)&     0 (    0)&     0 (    0)\\
10.50$-$11.00 &  4773 (    0) &     0 (    0)&     0 (    0)&     0 (    0)&     0 (    0)&     0 (    0)&     0 (    0)\\
\hline
Total & 31757 ( 6383) & 22604 ( 4512)& 13309 ( 2706)&  7853 ( 1618)&  4939 ( 1040)&   975 (  184)&   192 (   36)\\

\hline\hline
    \end{tabular*}
  \end{center}
\end{table*}

\begin{table*}
\caption{Same as Table 3, but for the deep survey.}
\begin{center}
\begin{tabular*}{\textwidth}{lrrrrrrr}
\hline\hline
&\multicolumn{6}{c}{Redshift}\\
$\log(L_{\rm IR}/$L$_{\odot})$ & 0.5$-$1 & 1$-$1.5 & 1.5$-$2.0 &2.0$-$2.5 & 2.5$-$3.0 & 3.0$-$4.0 & $>$ 4.0 \\
\hline%
13.00$-$       &     0 (    0) &     2 (    1)&     9 (    4)&    25 (   10)&    46 (   19)&    23 (   10)&    52 (   18)\\
12.50$-$13.00 &     7 (    3) &    65 (   23)&   159 (   58)&   250 (   93)&   308 (  122)&   165 (   70)&    73 (   25)\\
12.25$-$12.50 &    32 (   11) &   182 (   63)&   301 (  107)&   351 (  131)&   356 (  144)&   150 (   42)&    46 (    0)\\
12.00$-$12.25 &   114 (   39) &   445 (  153)&   552 (  196)&   540 (  204)&   457 (  167)&   106 (   10)&     1 (    0)\\
11.75$-$12.00 &   297 (  101) &   819 (  281)&   813 (  290)&   614 (  182)&   338 (   10)&    24 (    0)&     0 (    0)\\
11.50$-$11.75 &   606 (  205) &  1225 (  420)&   933 (  269)&   422 (    6)&    49 (    0)&     0 (    0)&     0 (    0)\\
11.00$-$11.50 &  2592 (  874) &  2754 (  567)&   826 (   27)&    48 (    0)&     0 (    0)&     0 (    0)&     0 (    0)\\
10.50$-$11.00 &  3574 (  804) &   466 (    0)&     0 (    0)&     0 (    0)&     0 (    0)&     0 (    0)&     0 (    0)\\
      $-$10.50 &  1031 (   20) &     0 (    0)&     0 (    0)&     0 (    0)&     0 (    0)&     0 (    0)&     0 (    0)\\
\hline
Total &  8254 ( 2057) &  5957 ( 1507)&  3593 (  951)&  2249 (  627)&  1554 (  462)&   467 (  133)&   172 (   44)\\
\hline\hline%
   \end{tabular*}
\end{center}
\label{tab:pahdeep}
\end{table*}

\begin{table*}
\caption{Numbers of the AGNs (AGN100\% and SF50\%+AGN50\%) expected to be detected in the SMI/LR wide survey. The values in the parentheses are the numbers of SF50\%+AGN50\% galaxies.}
%    \small
\begin{center}
\begin{tabular*}{\textwidth}{lrrrrrrr}
\hline\hline
&\multicolumn{6}{c}{Redshift}\\
$\log(L_{\rm IR}/$L$_{\odot})$ & 0$-$1 & 1$-$1.5 & 1.5$-$2.0 & 2.0$-$2.5 & 2.5$-$3.0 & 3.0$-$4.0 & $>$ 4.0 \\
\hline%
13.00$-$       & 2 (1)& 15 (9)& 60 (38)& 157 (101)& 310 (193)& 185 (104)& 310 (110)\\
12.50$-$13.00 & 37 (28)& 276 (231)& 710 (576)& 1240 (934)& 1639 (1054)& 596 (195)& 146 (5)\\
12.25$-$12.50 & 130 (114)& 701 (631)& 1264 (1068)& 1270 (852)& 694 (45)& 84 (0)& 0 (0)\\
12.00$-$12.25 & 434 (401)& 1661 (1527)& 1699 (1341)& 551 (34)& 89 (0)& 0 (0)& 0 (0)\\
11.75$-$12.00 & 1114 (1054)& 2703 (2472)& 576 (155)& 30 (0)& 0 (0)& 0 (0)& 0 (0)\\
11.50$-$11.75 & 2293 (2195)& 1664 (1347)& 56 (0)& 0 (0)& 0 (0)& 0 (0)& 0 (0)\\
11.00$-$11.50 & 6161 (5863)& 143 (0)& 0 (0)& 0 (0)& 0 (0)& 0 (0)& 0 (0)\\
10.50$-$11.00 & 4346 (4206)& 0 (0)& 0 (0)& 0 (0)& 0 (0)& 0 (0)& 0 (0)\\
     $-$10.50 & 2770 (2724)& 0 (0)& 0 (0)& 0 (0)& 0 (0)& 0 (0)& 0 (0)\\
\hline
Total & 17288 (16585)& 7162 (6217)& 4365 (3178)& 3248 (1921)& 2732 (1293)& 864 (299)& 456 (116)\\
\hline\hline
\end{tabular*}
\end{center}
\label{tab:agnwide}
\end{table*}

\begin{table*}
\caption{Same as Table 5, but for the deep survey.}
\begin{center}
\begin{tabular*}{\textwidth}{lrrrrrrr}
\hline\hline
&\multicolumn{6}{c}{Redshift}\\
$\log(L_{\rm IR}/$L$_{\odot})$ & 0$-$1 & 1$-$1.5 & 1.5$-$2.0 & 2.0$-$2.5 & 2.5$-$3.0 & 3.0$-$4.0 & $>$ 4.0 \\
\hline%
13.00$-$       & 0 (0)& 2 (1)& 6 (4)& 16 (10)& 31 (19)& 19 (10)& 71 (36)\\
12.50$-$13.00 & 4 (3)& 28 (23)& 71 (58)& 124 (93)& 181 (122)& 116 (70)& 212 (92)\\
12.25$-$12.50 & 13 (11)& 70 (63)& 126 (107)& 173 (131)& 219 (144)& 144 (77)& 84 (10)\\
12.00$-$12.25 & 43 (40)& 166 (153)& 232 (196)& 278 (204)& 326 (199)& 141 (25)& 22 (0)\\
11.75$-$12.00 & 111 (105)& 304 (281)& 348 (290)& 346 (232)& 221 (35)& 52 (0)& 0 (0)\\
11.50$-$11.75 & 229 (219)& 455 (420)& 407 (322)& 169 (26)& 52 (0)& 0 (0)& 0 (0)\\
11.00$-$11.50 & 1022 (988)& 982 (873)& 199 (73)& 29 (0)& 0 (0)& 0 (0)& 0 (0)\\
10.50$-$11.00 & 1551 (1501)& 66 (31)& 0 (0)& 0 (0)& 0 (0)& 0 (0)& 0 (0)\\
     $-$10.50 & 1377 (1347)& 0 (0)& 0 (0)& 0 (0)& 0 (0)& 0 (0)& 0 (0)\\
\hline
Total & 4351 (4215)& 2072 (1844)& 1389 (1049)& 1134 (697)& 1031 (519)& 472 (183)& 388 (139)\\
\hline\hline
\end{tabular*}
\end{center}
\label{tab:agndeep}
\end{table*}

\begin{table*}
\caption{Numbers of the AGNs (AGN100\% and SF50\%+AGN50\%) expected to
  be detected with SMI/CAM in the wide survey. The values in the
  parentheses are those with SMI/CAM in the deep survey.}
\footnotesize
\begin{center}
\begin{tabular*}{\textwidth}{lrrrrrrr}
\hline\hline
&\multicolumn{6}{c}{Redshift}\\
$\log(L_{\rm IR}/$L$_{\odot})$ & 0$-$1 & 1$-$1.5 & 1.5$-$2.0 & 2.0$-$2.5 & 2.5$-$3.0 & 3.0$-$4.0 & $>$ 4.0 \\
\hline
13.00$-$       & 2 (0)& 15 (2)& 60 (6)& 157 (16)& 310 (31)& 185 (19)& 782 (78)\\
12.50$-$13.00 & 37 (4)& 276 (28)& 710 (71)& 1240 (124)& 1808 (181)& 1161 (116)& 4901 (490)\\
12.25$-$12.50 & 130 (13)& 701 (70)& 1264 (126)& 1727 (173)& 2195 (219)& 1537 (154)& 6399 (649)\\
12.00$-$12.25 & 434 (43)& 1661 (166)& 2317 (232)& 2775 (278)& 3360 (336)& 2616 (262)& 9435 (1105)\\
11.75$-$12.00 & 1114 (111)& 3035 (304)& 3484 (348)& 3867 (387)& 4618 (462)& 4081 (408)& 10511 (1722)\\
11.50$-$11.75 & 2293 (229)& 4553 (455)& 4592 (459)& 4916 (492)& 5885 (589)& 5889 (589)& 8654 (2270)\\
11.00$-$11.50 & 10216 (1022)& 13184 (1318)& 12090 (1209)& 12708 (1271)& 14626 (1534)& 12394 (1776)& 4699 (4109)\\
10.50$-$11.00 & 19871 (1987)& 17380 (1738)& 13916 (1548)& 9018 (1630)& 4667 (1962)& 1005 (2025)& 0 (1073)\\
     $-$10.50 & 69788 (11097)& 14358 (4171)& 2503 (2280)& 117 (1427)& 0 (841)& 0 (278)& 0 (0)\\
\hline
Total & 103886 (14506)& 55164 (8251)& 40935 (6279)& 36526 (5795)& 37469 (6154)& 28868 (5626)& 45382 (11497)\\
\hline\hline
\end{tabular*}
\end{center}
\label{tab:agnwidecam}
\end{table*}

\subsection{Characterization of PAHs}
We have created simulated SMI/LR PAH spectra in order to confirm the result
obtained in the previous subsection, and also the applicability of the
PAH features to determine the redshift and to characterize the
PAHs in a galaxy. The model spectrum was taken from that of typical
Galactic diffuse PAH emission \citep{draine07} plus an M82-like continuum approximated by a power-law component. We assumed galaxies at redshift $z=3$ with three levels of total IR luminosities, $L_{\rm IR} = 1\times10^{12}$, $3\times10^{12}$, and $1\times10^{13}$~L$_{\odot}$. Then $L_{\rm PAH7.7}$ is determined and fixed according to the relation in Table~\ref{tab:strength}, while $L_{\rm PAH6.2}$ and $L_{\rm PAH8.6}$ are allowed to vary by a factor of 3 relative to $L_{\rm PAH7.7}$. The latter reflects a possible systematic difference in the properties of PAHs in galaxies at high-$z$ when compared to those in the nearby universe as well as their intrinsic variations from galaxy to galaxy. 

\begin{figure*}[htbp]
\begin{center}
  \includegraphics[width=\columnwidth]{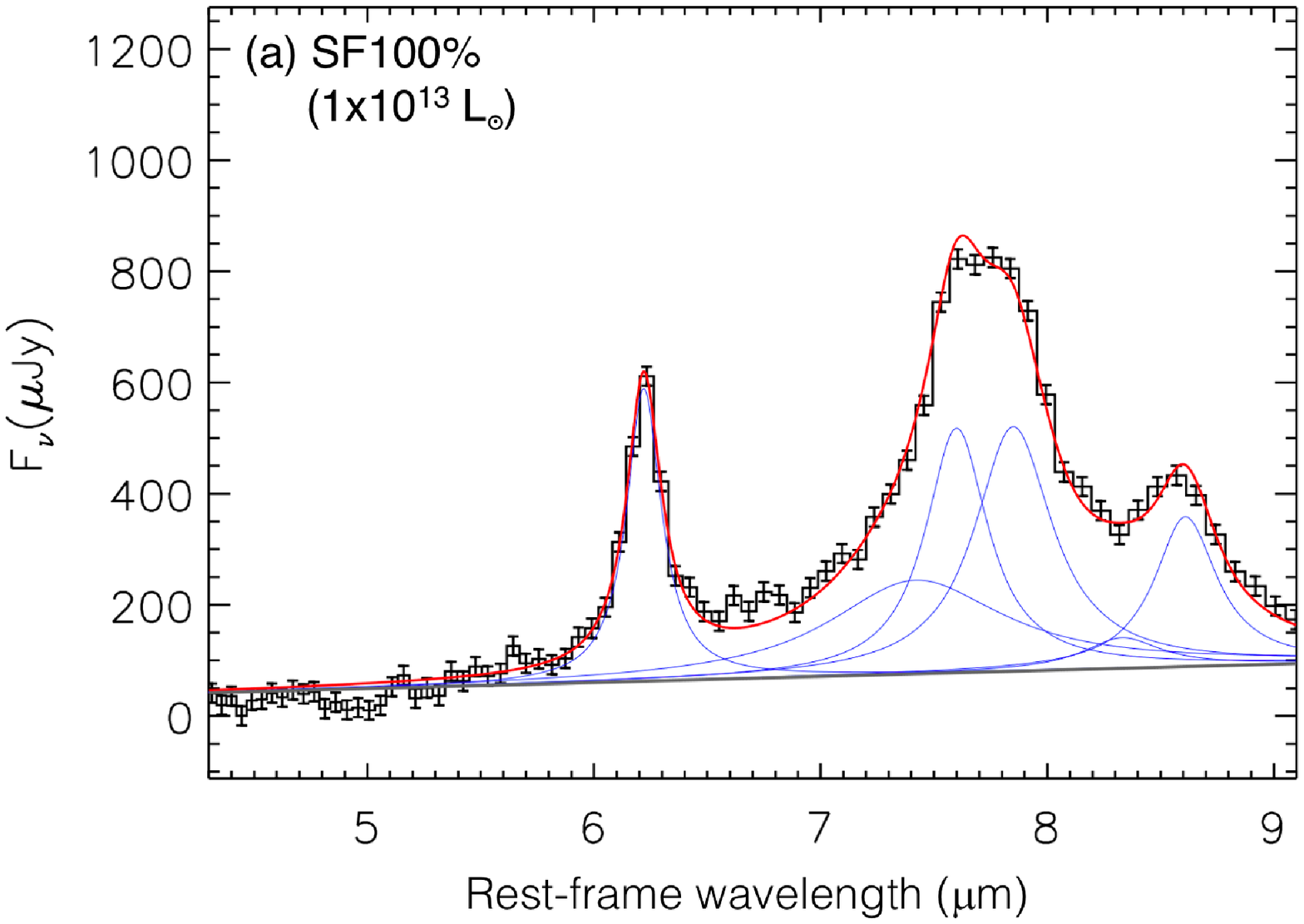}
  \includegraphics[width=\columnwidth]{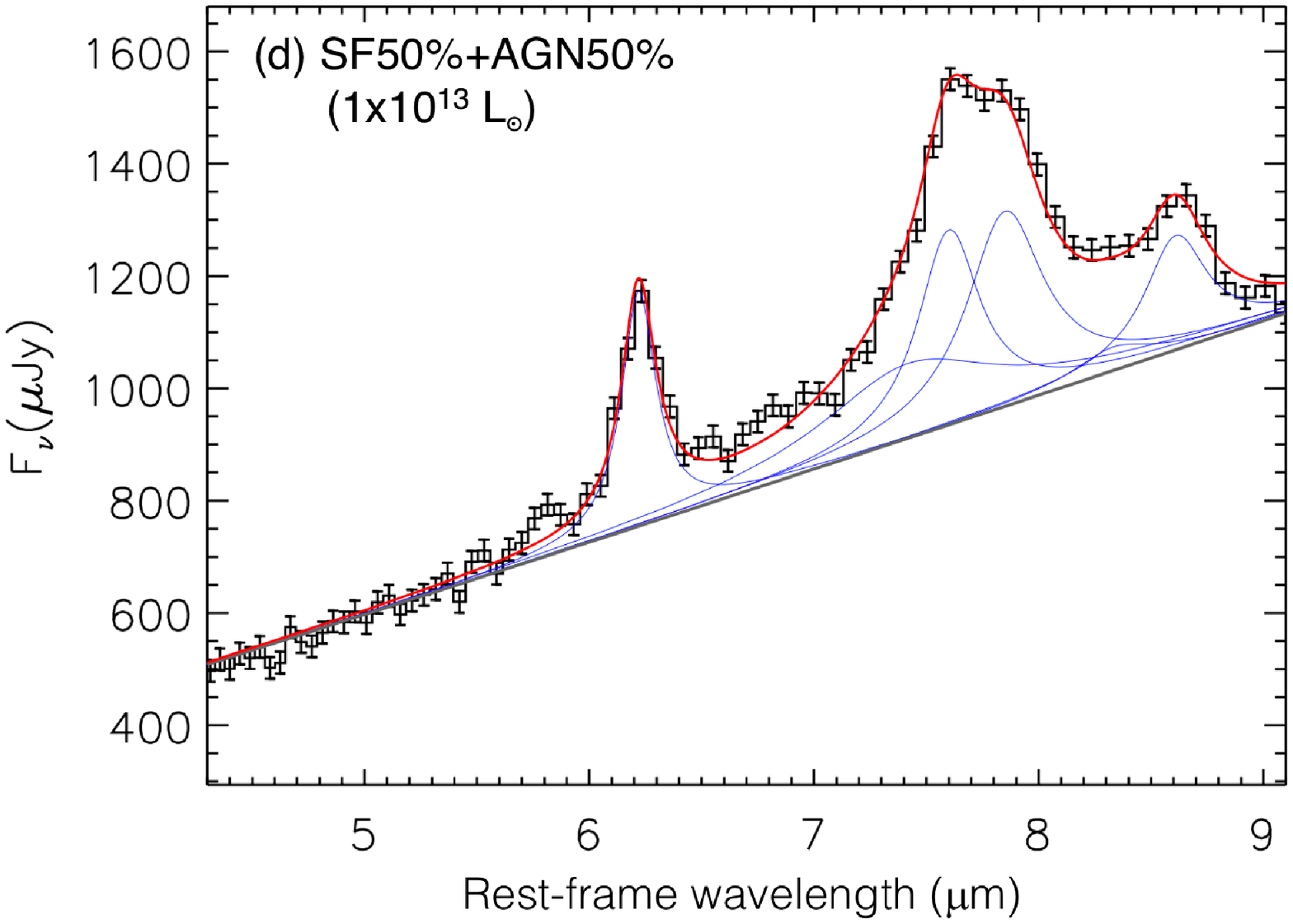} %
  \includegraphics[width=\columnwidth]{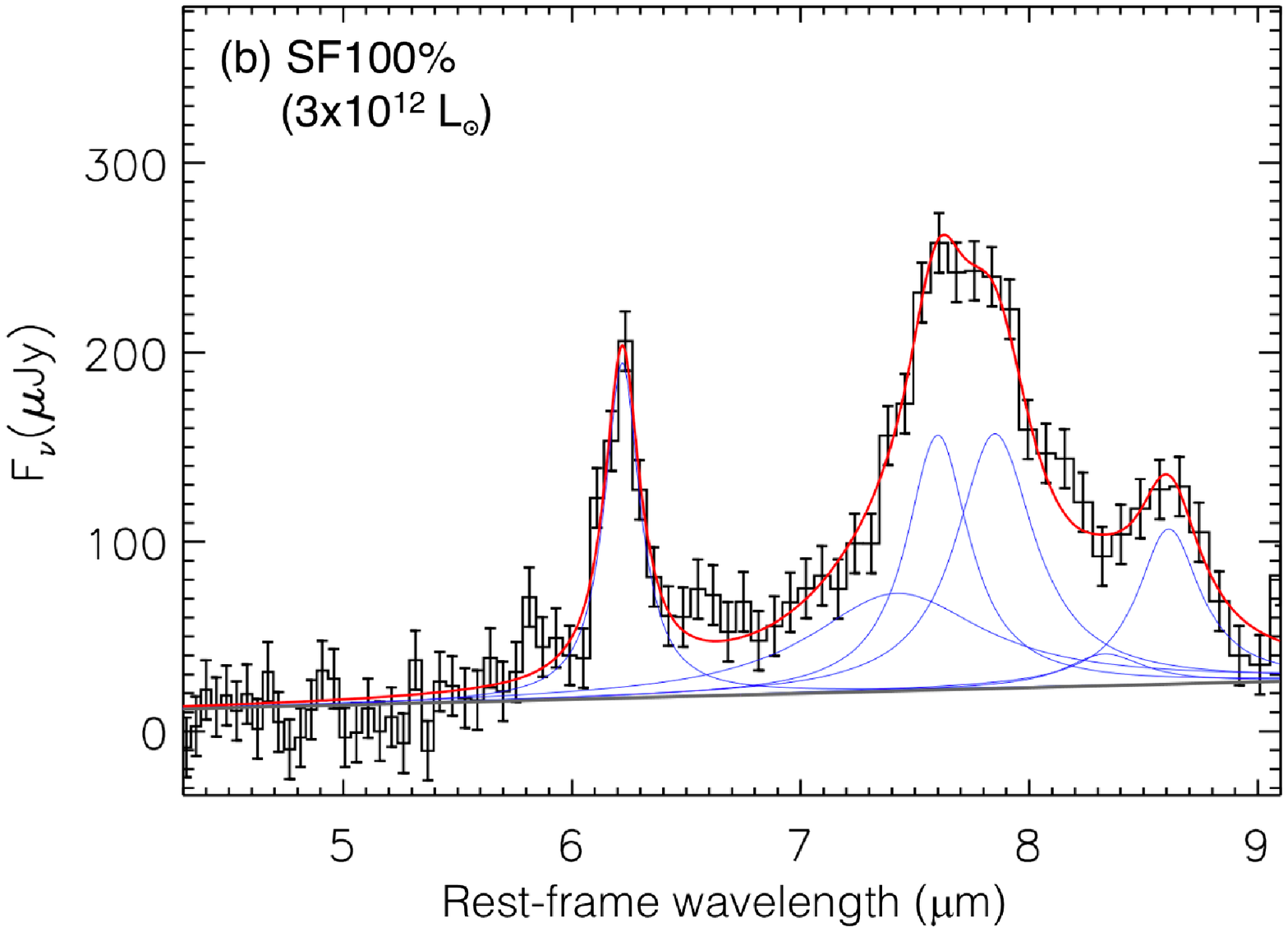}
  \includegraphics[width=\columnwidth]{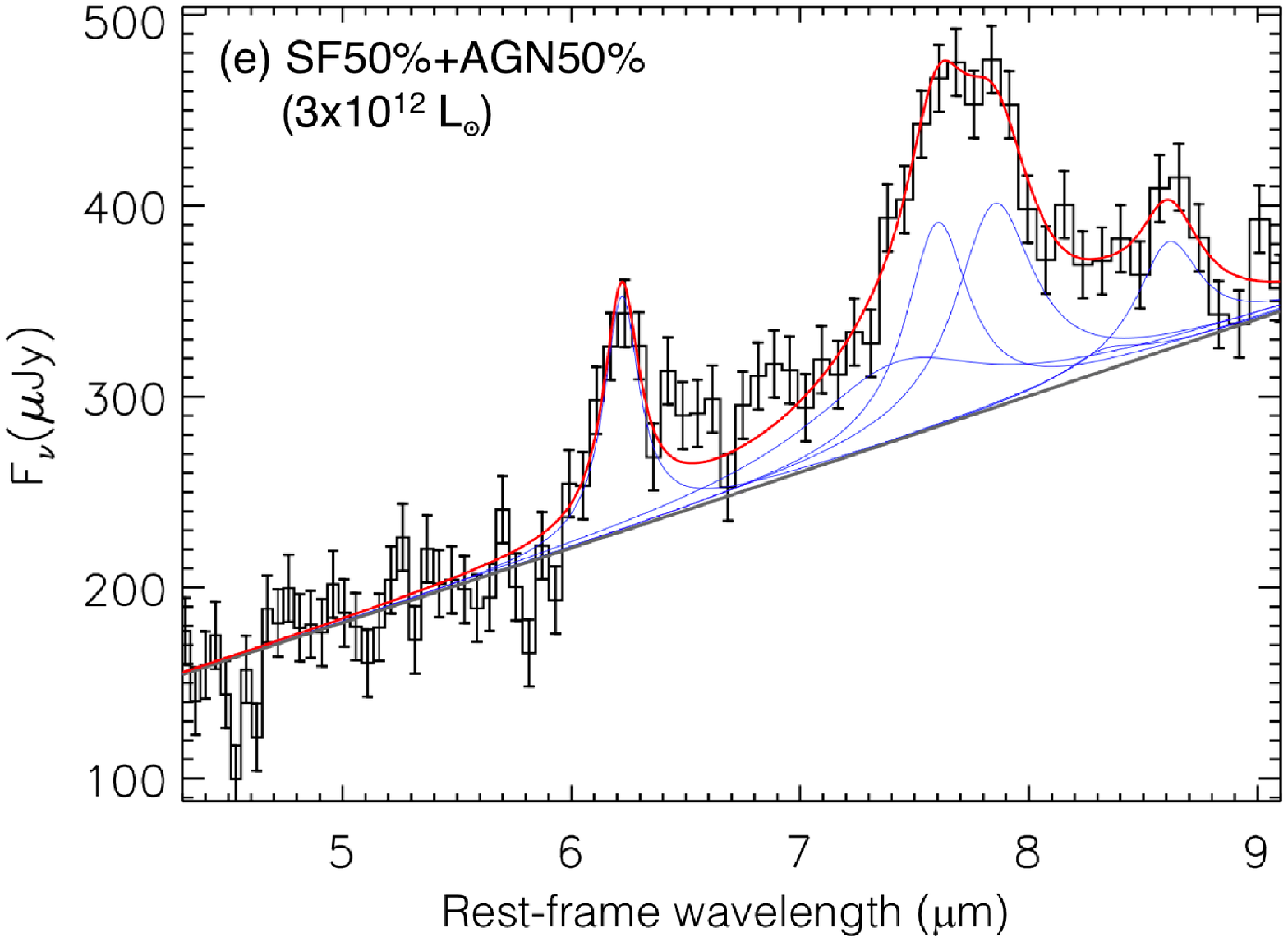} %
  \includegraphics[width=\columnwidth]{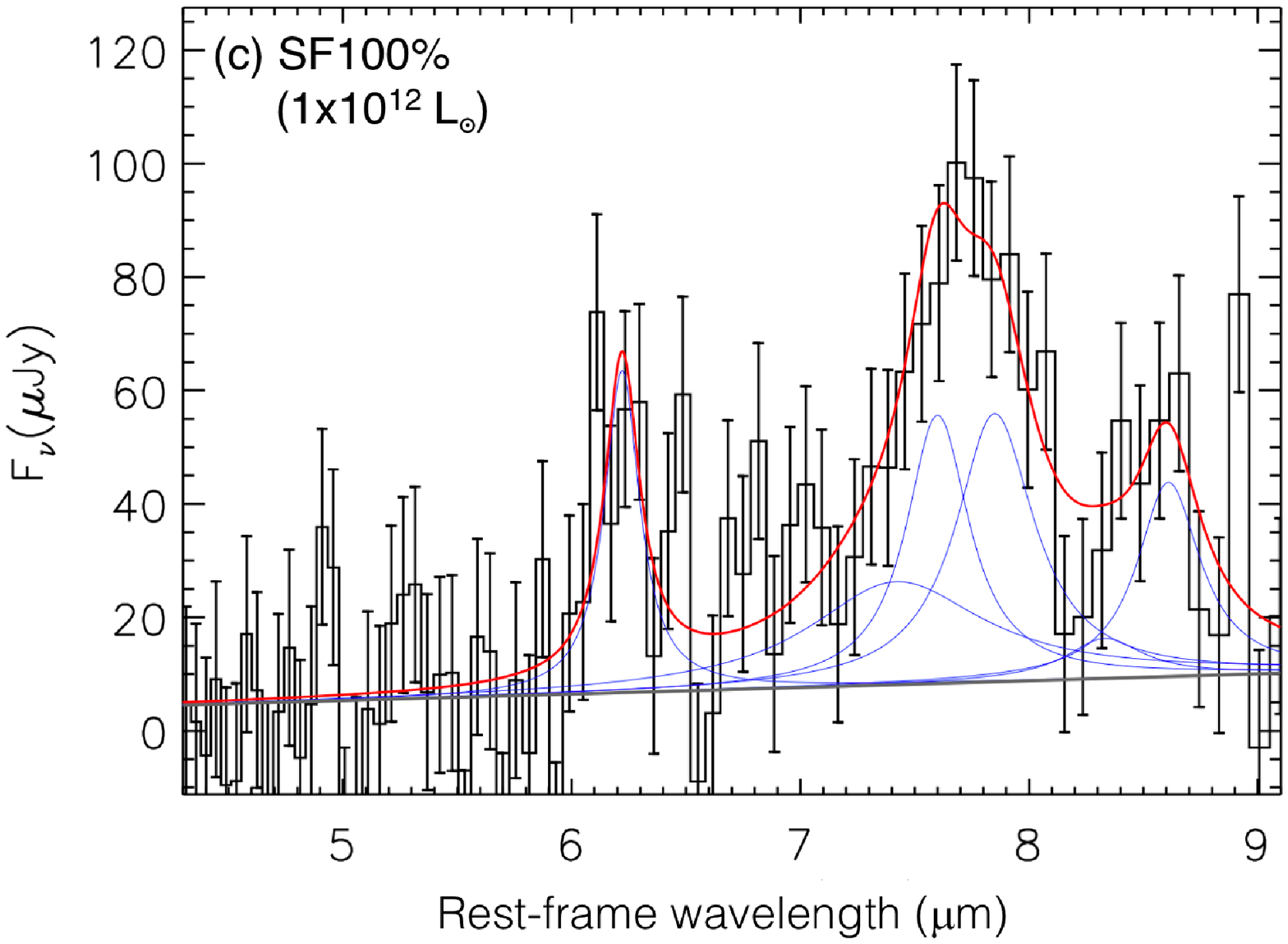}
  \includegraphics[width=\columnwidth]{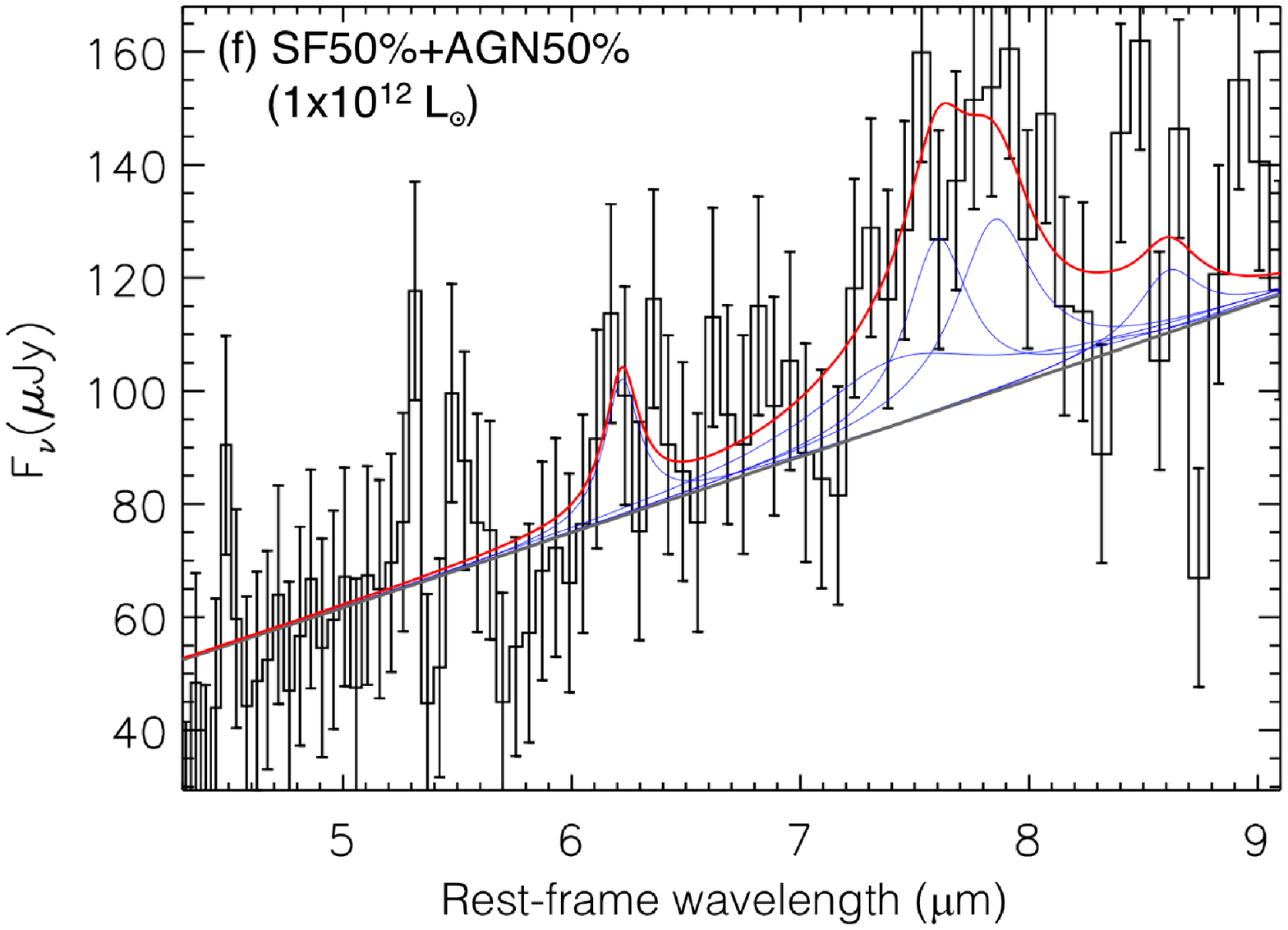}
  \caption{Simulated SMI/LR spectra of a galaxy at $z=3$ for the deep survey, SF100\% on the left and SF50\%+AGN50\% on the right-hand side with $L_{\rm IR}$ denoted in each panel. Solid curves indicate best-fit results with PAHFIT.}
  \label{fig6}
\end{center}
\end{figure*}

Figure~\ref{fig6} shows examples of the simulated spectra of SF100\% and
SF50\%+AGN50\% galaxies at $z=3$ for the SMI/LR deep survey. As for the
AGN continuum, we utilized an AGN spectral template \citep{polletta07}
and scaled the amplitude so that the IR luminosity integrated from 8 to
1000~$\mu$m reaches a specified value (i.e., $0.5\times L_{\rm
IR}$). A Nyquist sampling for $R = 50$ resolution was adopted, adding white noise with amplitude based on the continuum sensitivity expected for the deep survey ($\sim$100~$\mu$Jy, $5\sigma$). In order to fit the spectra, we used PAHFIT \citep{smith07}, assuming a power-law continuum and an extinction curve with a screen configuration \citep{kemper04}.
Free parameters are the normalizations of the PAH features, the normalization and the index of the power-law continuum, and the extinction. PAHFIT assumes that the PAH 7.7~$\mu$m and 8.6~$\mu$m features are the complexes consisting of 7.4, 7.6, and 7.8~$\mu$m sub-features and 8.3 and 8.6~$\mu$m sub-features, respectively. In both generating and fitting the simulated spectra, we fixed the relative intensity ratios among the sub-features at typical values for each complex \citep{draine07}, but allowed the PAH 8.6~$\mu$m feature to vary with respect to the PAH 7.7~$\mu$m feature.

\begin{figure*}[htbp]
\begin{center}
  \includegraphics[width=\columnwidth]{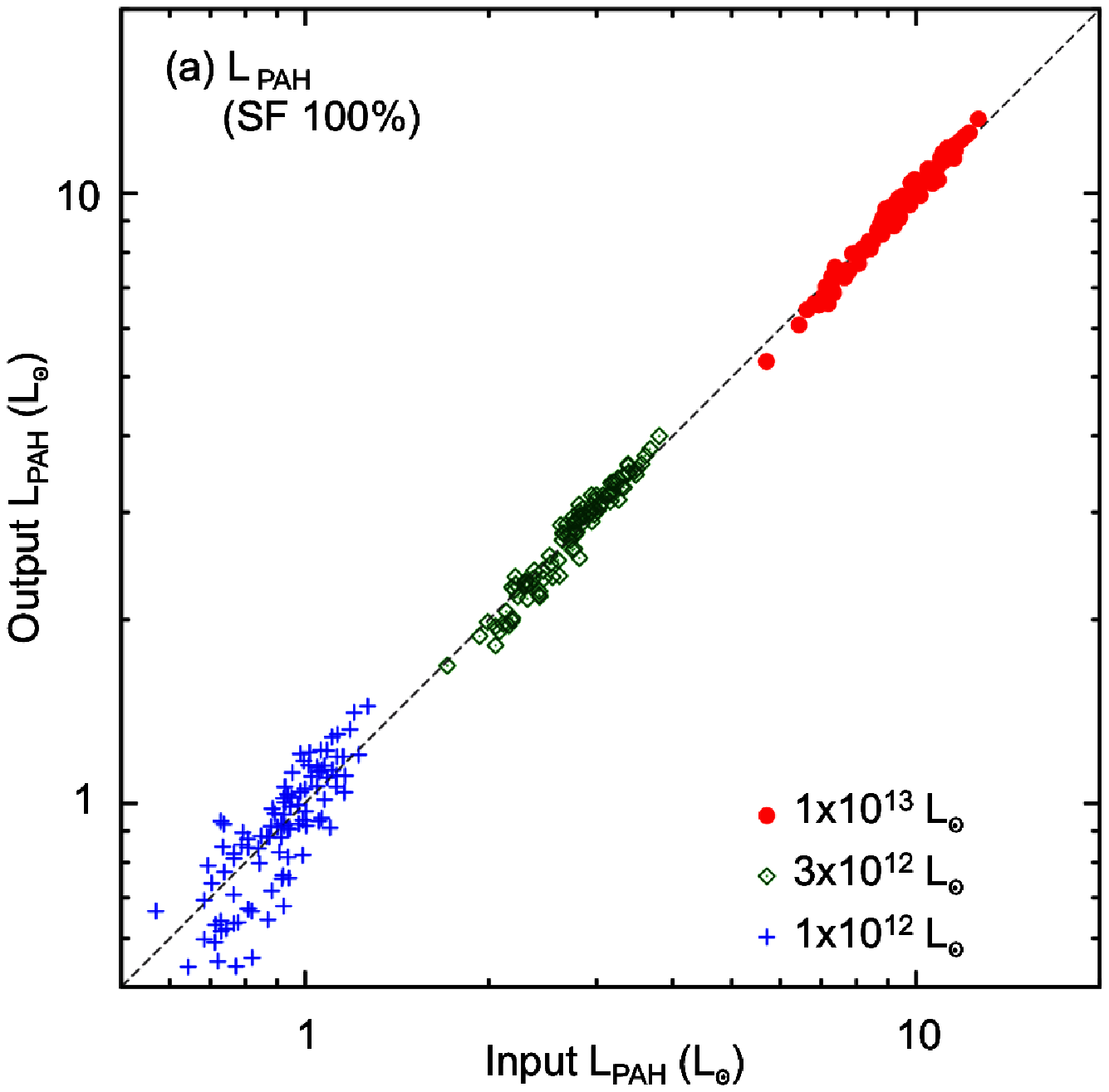}
  \includegraphics[width=\columnwidth]{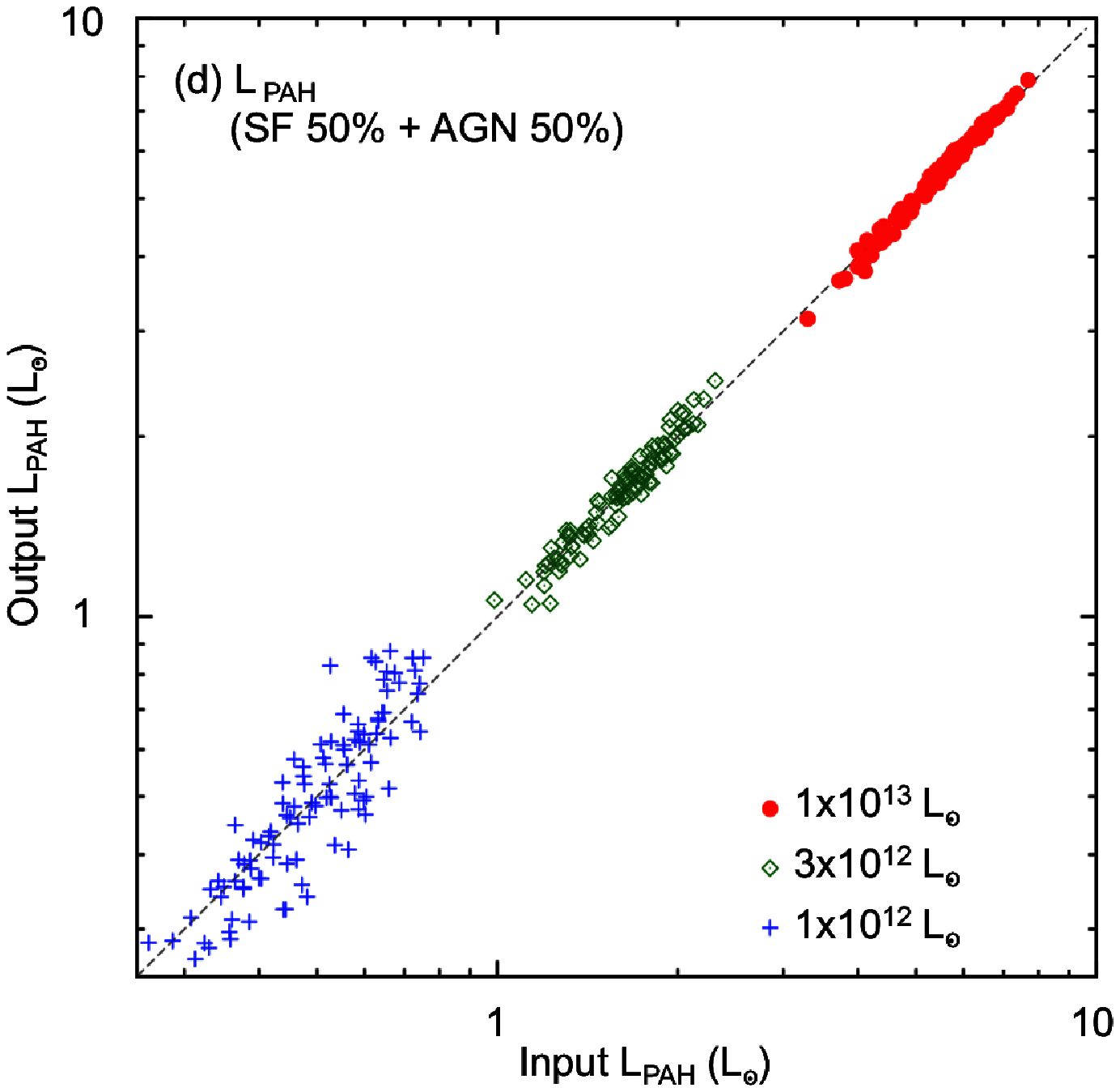}\\

  \includegraphics[width=\columnwidth]{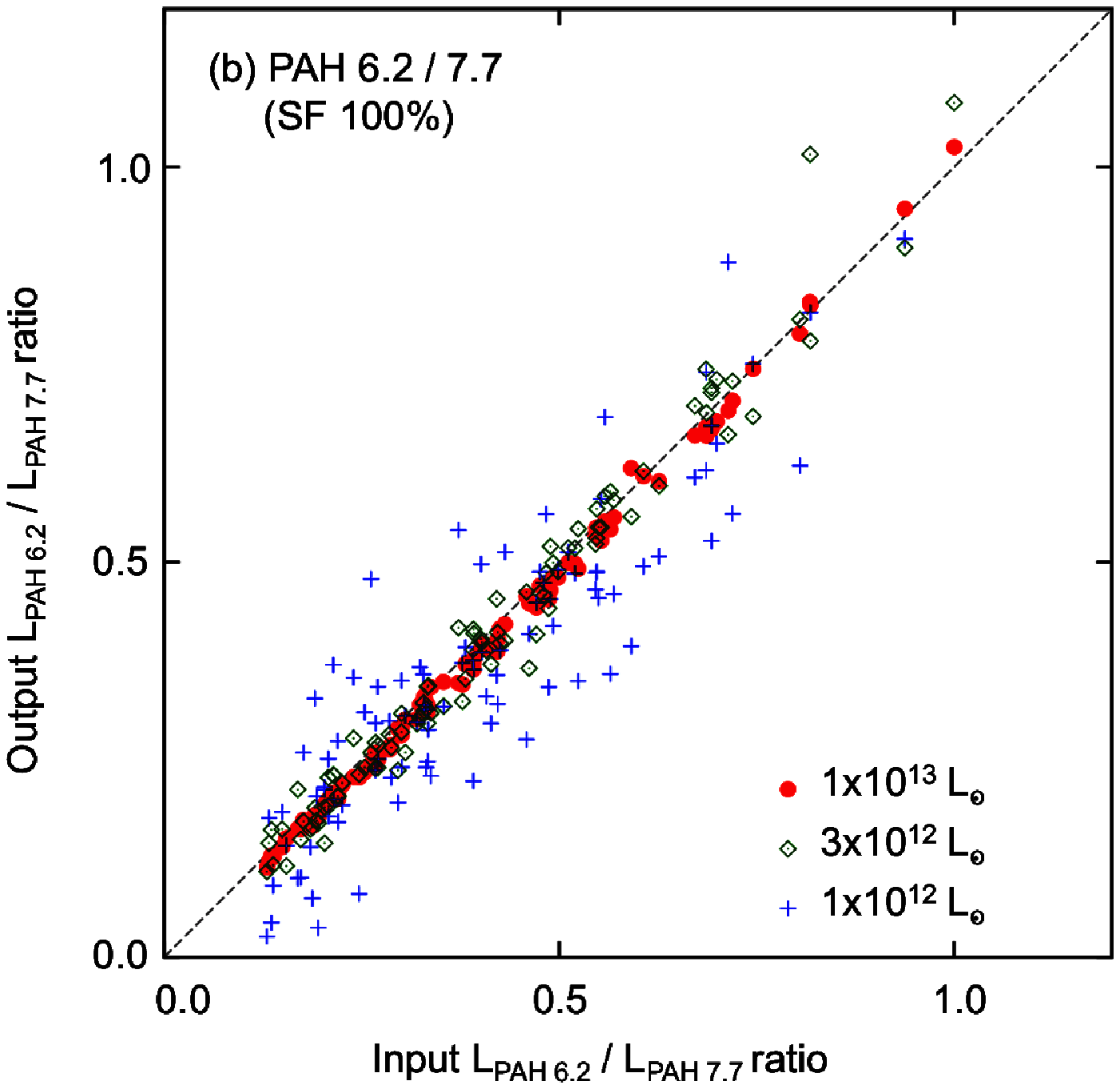}
  \includegraphics[width=\columnwidth]{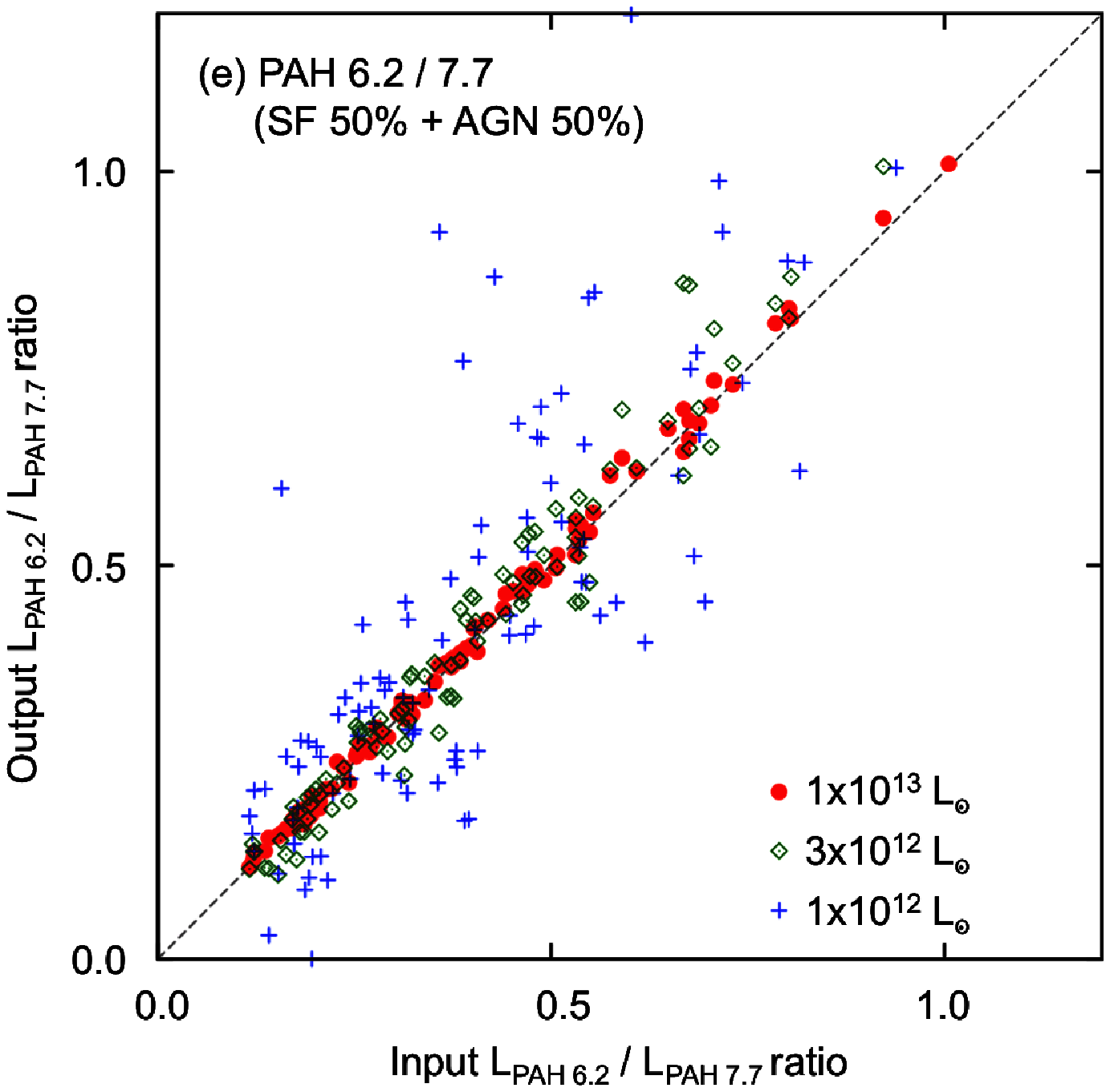}\\

  \includegraphics[width=\columnwidth]{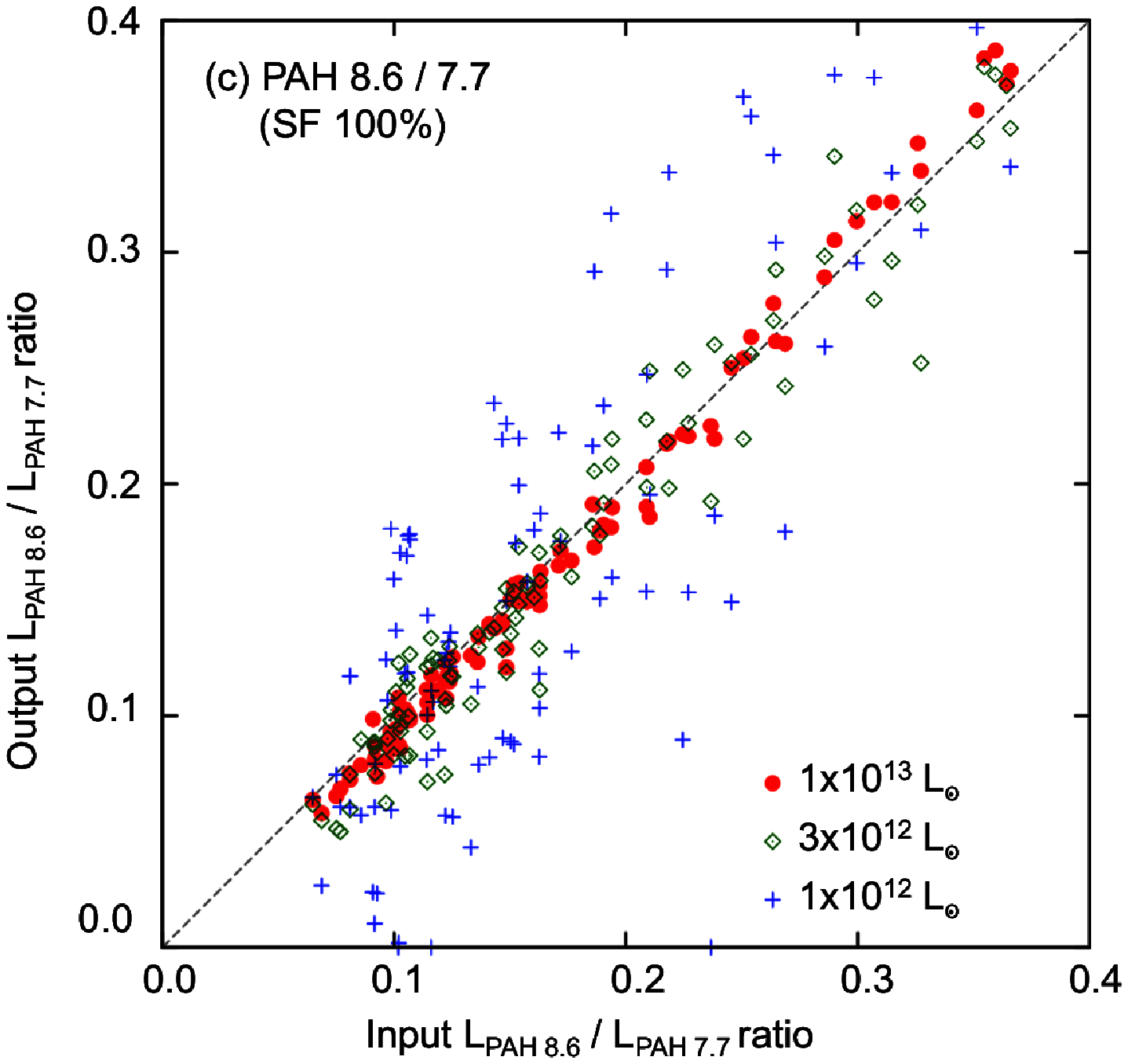}
  \includegraphics[width=\columnwidth]{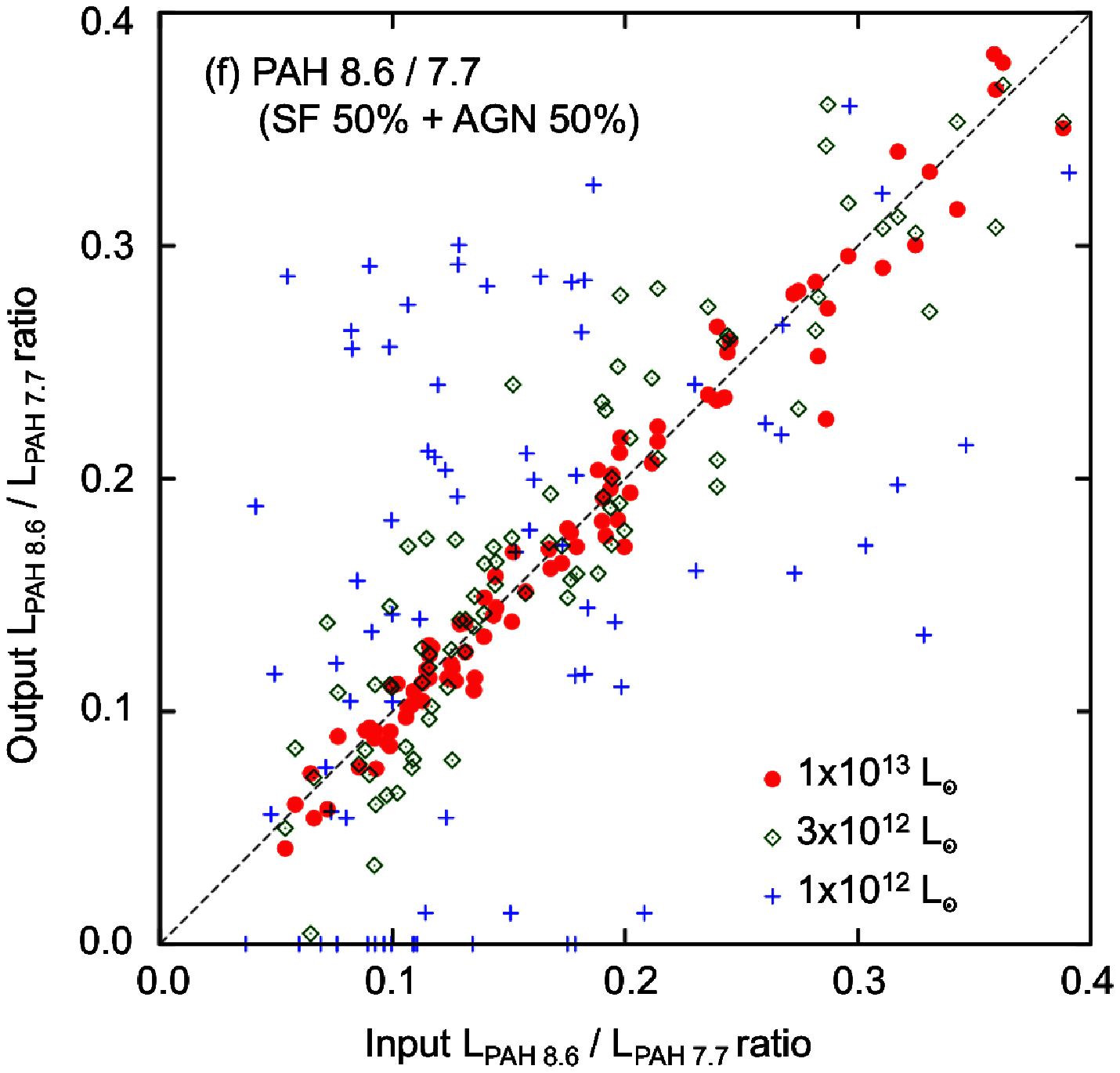}
  \caption{Correlation plots between output (measured) and input (simulated) values of (a, d) $L_{\rm PAH}$, (b, e) $L_{\rm PAH6.2}/L_{\rm PAH7.7}$ and (c, f) $L_{\rm PAH8.6}/L_{\rm PAH7.7}$ of galaxies at $z=3$ for the deep survey, SF100\% on the left and SF50\%+AGN50\% on the right-hand side. The dashed line in each panel corresponds to $y=x$.}
\label{fig7}
\end{center}
\end{figure*}

First we take results of spectral fitting for SF100\% galaxies. We confirm that the redshift is determined with the accuracies of 2\%, 0.7\%, and 0.3\% for galaxies with $L_{\rm IR}$ of $1\times10^{12}$, $3\times10^{12}$, and $1\times10^{13}$~L$_{\odot}$, respectively. Figure~\ref{fig7}a shows a correlation plot between the output (measured) and input (simulated) values of $L_{\rm PAH}$ ($\equiv L_{\rm PAH6.2}+L_{\rm PAH7.7}+L_{\rm PAH8.6}$). From the figure, we find that $L_{\rm PAH}$ is determined with the accuracies of 13\% (bias: 0.8\%), 5\% (1\%), and 3\% (0.3\%) for galaxies with $L_{\rm IR}$ of $1\times10^{12}$, $3\times10^{12}$, and $1\times10^{13}$~L$_{\odot}$, respectively; here and hereafter the accuracy is defined by the standard deviation of the absolute values of the differences between the input and output divided by the input values, while the bias is the systematic difference between the input and output values.
Hence the SMI/LR surveys can estimate the redshift very precisely, and
also measure $L_{\rm PAH}$ of so-called main-sequence galaxies at $z=3$
($\sim 1\times 10^{12}$~L$_{\odot}$). Figures~\ref{fig7}b and \ref{fig7}c show correlation plots between the input and output values of $L_{\rm PAH6.2}/L_{\rm PAH7.7}$ and $L_{\rm PAH8.6}/L_{\rm PAH7.7}$, respectively. From the figure, we find that $L_{\rm PAH6.2}/L_{\rm PAH7.7}$ is determined with the accuracies of 31\% (bias: $-$0.8\%), 9\% (0.2\%), and 4\% ($-$0.3\%), while $L_{\rm PAH8.6}/L_{\rm PAH7.7}$ is determined with 47\% (4.3\%), 13\% (1\%), and 6\% ($-$1\%) for galaxies with $L_{\rm IR}$ of $1\times10^{12}$, $3\times10^{12}$, and $1\times10^{13}$~L$_{\odot}$, respectively. Thus the result also demonstrates applicability of those features to characterize PAHs in galaxies at $z\sim 3$. 

\begin{figure}[htbp]
\begin{center}
  \includegraphics[width=\columnwidth]{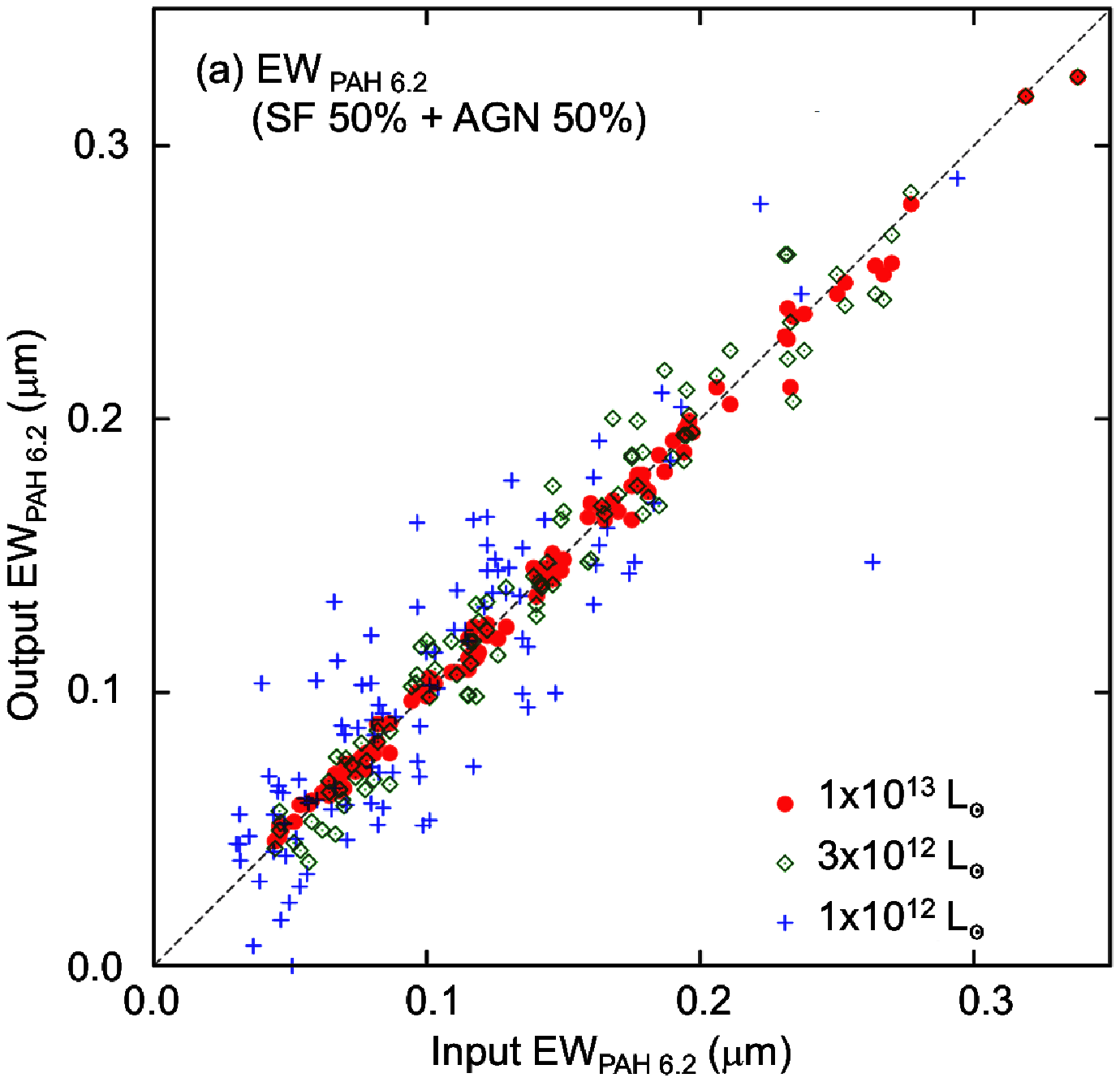}
  \includegraphics[width=\columnwidth]{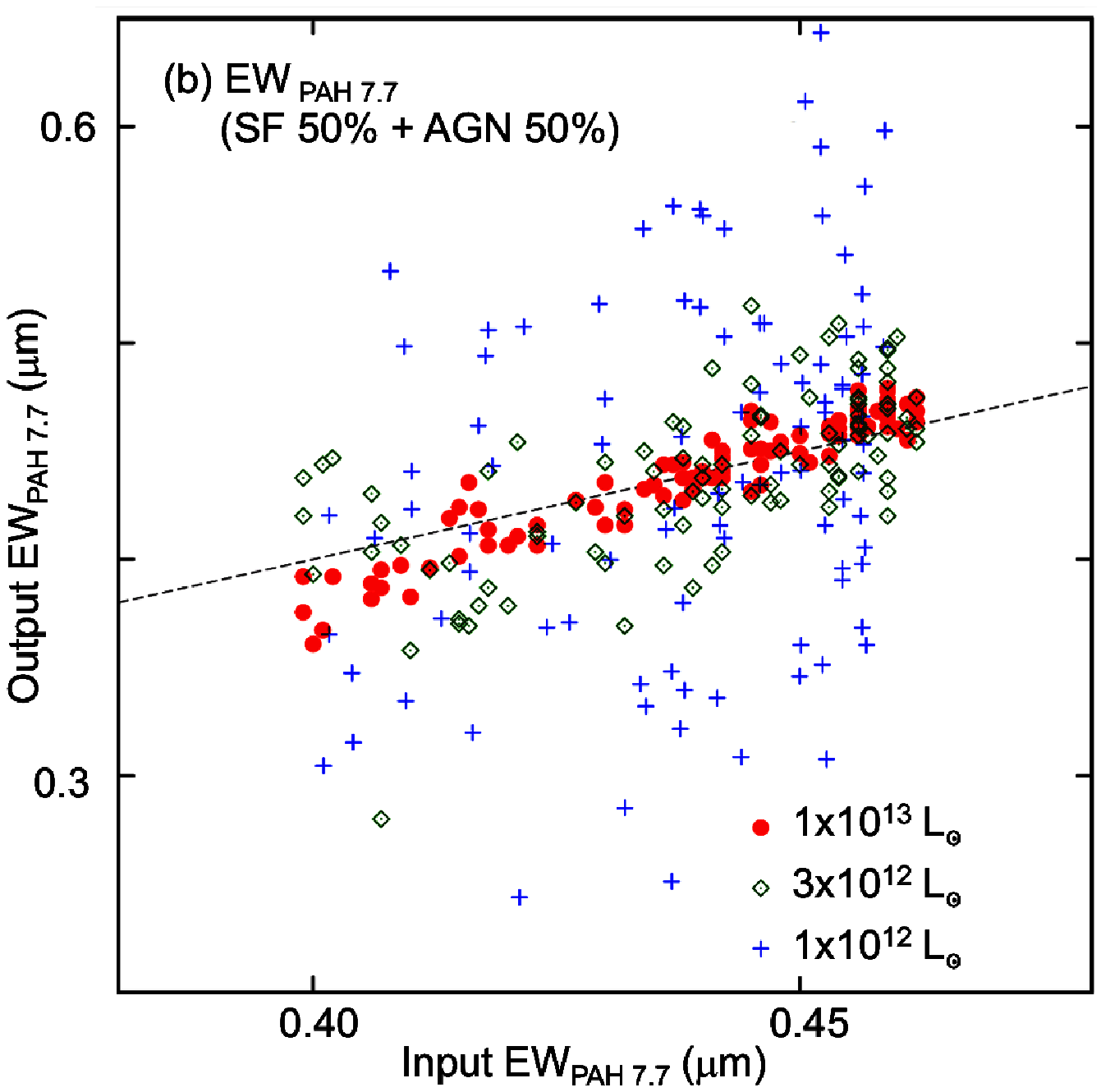}
  \caption{Correlation plots between output (measured) and input (simulated) values of (a) the PAH 6.2~$\mu$m and (b) 7.7~$\mu$m equivalent widths of SF50\%+AGN50\% galaxies at $z=3$ for the deep survey. The dashed line in each panel corresponds to $y=x$.}
\label{fig_eqw}
\end{center}
\end{figure}

For SF50\%+AGN50\% galaxies, the above accuracies are degraded to 3\%,
0.8\%, and 0.6\% for the redshift, 19\% (bias: 2\%), 8\% (1\%), and 3\% 
(0.5\%) for $L_{\rm PAH}$, 43\% (9\%), 12\% ($-$2\%), and 4\% ($-$0.4\%) for $L_{\rm PAH6.2}/L_{\rm PAH7.7}$,
85\% (18\%), 24\% (8\%), and 8\% (0.5\%) for $L_{\rm
PAH8.6}/L_{\rm PAH7.7}$ for galaxies with $L_{\rm IR}$ of
$1\times10^{12}$, $3\times10^{12}$, and $1\times10^{13}$~L$_{\odot}$,
respectively. The degradation is caused by the decrease in $L_{\rm
PAH}/L_{\rm IR}$ for SF50\%+AGN50\% galaxies by a factor of 2 from that
for SF100\% galaxies. Figure~\ref{fig_eqw} shows a correlation plot for
the PAH equivalent width values used to estimate the
star-formation contribution to the total $L_{\rm IR}$ of a galaxy. We
find that the accuracies of the equivalent widths are 34\% (bias:
$-$4\%), 10\% ($-$1\%), and 4\% ($-$0.1\%) for the PAH 6.2~$\mu$m
and 21\% (1\%), 7\% (1\%), and 3\% (0.2\%) for the PAH
7.7~$\mu$m features. Hence we can estimate the relative proportions of 
the AGN and star-formation contributions to $L_{\rm IR}$ in galaxies down to a luminosity level
of $\sim 1\times10^{12}$~L$_{\odot}$, although we need $L_{\rm IR} >
3\times10^{12}$~L$_{\odot}$ to characterize the PAH emission bands. 

In the above cases, three PAH features are used for the redshift determination. Here we estimate the degradation of their accuracies as the number of PAH features decreases. In the SF100\% case, if we use only two features (6.2 and 7.7~$\mu$m), the accuracies are degraded to 2\%, 0.9\%, and 0.3\% for $L_{\rm IR}$ = $1\times10^{12}$, $3\times10^{12}$, and $1\times10^{13}$~L$_\odot$, respectively. If we use only one feature (6.2~$\mu$m), they are 3\%, 1\% and 0.5\%. In the SF50\%+AGN50\% case, the accuracies are degraded to 5, 1, and 0.7\% in the case of two features, while they are 10, 2, and 0.8\% in the case of one feature. Taking these accuracies and the result in Figure~\ref{fig:fraction} into account, we estimate that the redshifts are determined with an accuracy of $\le 2$\% for 84\% of the SF100\% galaxies and 68\% of the SF50\%+AGN50\% galaxies at $z=2$--$4$.

\subsection{Synergies with SAFARI and future large facilities}

The SMI blind surveys will provide us with unbiased, uniform, and
statistically significant numbers of samples for follow-up spectroscopy
with the SPICA far-IR instrument, SAFARI (and also SMI/MR or /HR, if
necessary). Since the field-of-view of SAFARI is rather narrow \citep{pastor16}, the integrated survey strategy is important for defining unbiased studies. The spectral datasets obtained with SMI/LR will enable us to
estimate redshifts, IR luminosities, and fractional AGN luminosities of
follow-up candidates, and thus to plan strategic observations with
SAFARI. Then SAFARI would provide information on the properties of
atomic and ionic gas in a galaxy based on fine-structure
line diagnostics (e.g., [S III] 18, 33~$\mu$m, [O IV] 26~$\mu$m,
[Si II] 35~$\mu$m, [O III] 52, 88~$\mu$m, and [N III] 57~$\mu$m)
\citep{spinoglio12, spinoglio17, fernandez16}, which is complementary to
the information on dust features provided by SMI (e.g., PAHs, silicates).  

\begin{figure*}[htbp]
\begin{center}
\includegraphics[width=1.5\columnwidth]{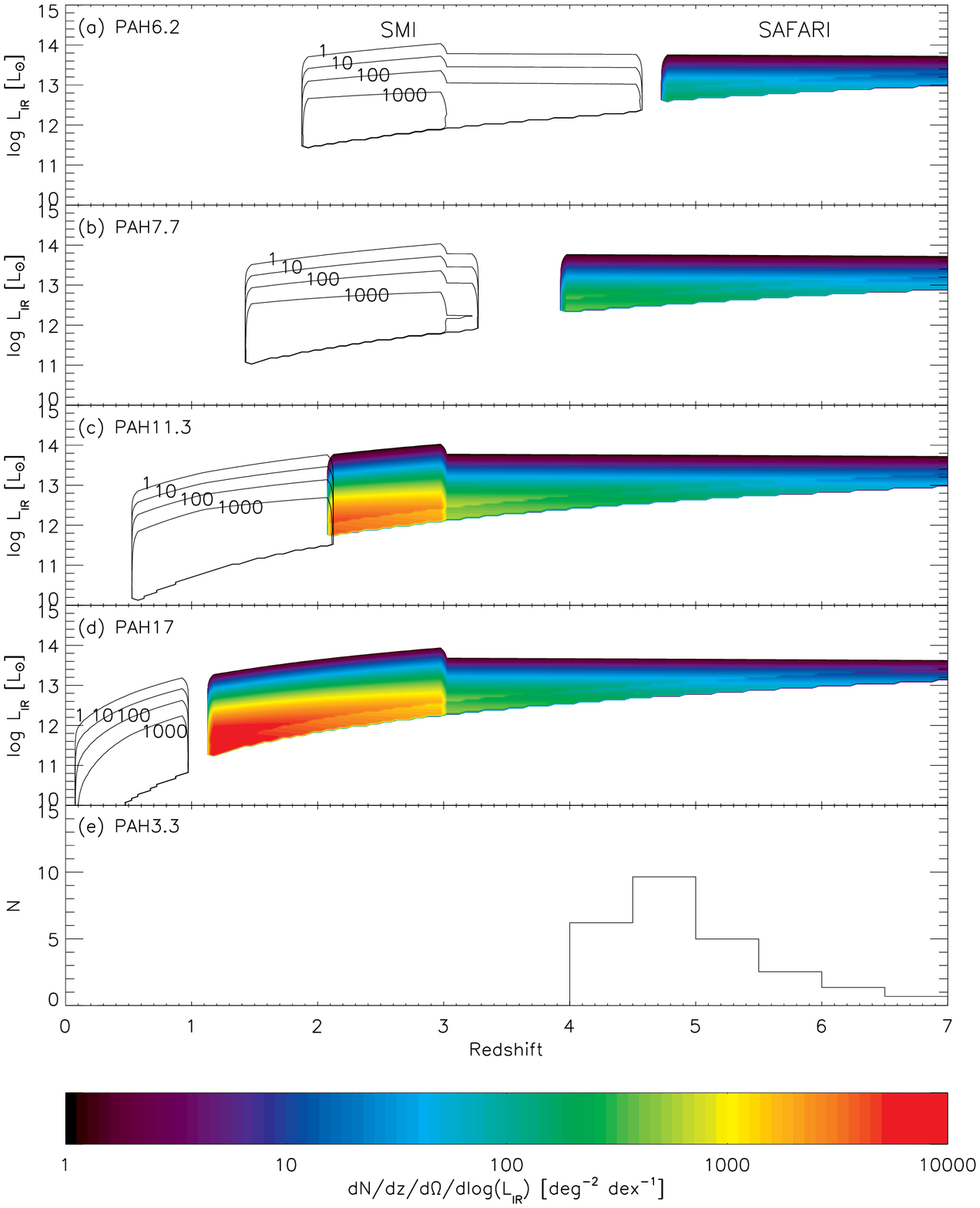}
\caption{Number densities of PAH galaxies per unit redshift, per sq. deg., and per unit $d\log(L_{\rm IR})$ detectable with SAFARI (colors) in 1-hour exposure using (a) the PAH 6.2, (b) 7.7, (c) 11.3, and (d) 17~$\mu$m features. The result of the SMI/LR deep survey is shown together with contours of 4 levels (1, 10, 100, 1000 per unit redshift, unit deg$^2$, and unit $d\log(L_{\rm IR})$). (e) Number of galaxies expected to be detected in the PAH 3.3~$\mu$m feature with SMI/LR in the deep survey.}
\label{fig9}
\end{center}
\end{figure*}

Beyond $z=1$, some of the PAH 6.2, 7.7, 11.3, and 17~$\mu$m features fall
outside of the SMI range, but will be covered by
SAFARI. Figures~\ref{fig9}a, b, c, and d visualize such complementarity
between SMI and SAFARI in detecting the PAH 6.2, 7.7, 11.3, and 17~$\mu$m features, respectively. The contour maps show the number
densities of the PAH galaxies expected to be detected by the SMI/LR deep
survey, while the color maps show those detectable with a SAFARI 1-hour
pointing spectroscopy. From the figure, we can confirm that the SMI and
SAFARI domains are connected smoothly to each other. Hence the
integrated SAFARI and SMI observations allow us not only to detect the
discrete PAH features but also to characterize the PAH emission in distant galaxies for the first time. 

SMI provides high-redshift samples beyond $z=4$ where the PAH 3.3 and
6.2~$\mu$m features are available at $z>4.3$ and $z=4$--$4.6$,
respectively, within the wavelength range of SMI/LR. For example, the expected number of galaxies beyond $z=4$ is $\sim 170$ for the deep survey (see Table \ref{tab:pahdeep}), most of which may be important targets to perform follow-up observations with SAFARI. In particular the PAH 3.3~$\mu$m feature (and possibly aliphatic sub-features at $3.4$--$3.6$~$\mu$m) could be a very powerful probe of high-redshift dusty galaxies, because there are no upper limits on the coverage of redshift practically, and its intrinsic bandwidth matches very well the instrumental spectral resolution ($R=50$--120) of SMI/LR. It should be noted that, at $z>5$, where the observed wavelength of the PAH 3.3~$\mu$m feature exceeds 20~$\mu$m, the $R=50$ continuum sensitivity of SMI/LR surpasses that of JWST/MIRI (e.g., 3 and 10 times higher at 20 and 28~$\mu$m, respectively, according to \citet{glasse15}). However the PAH 3.3~$\mu$m feature is relatively weak as compared to the other PAH features (see Table \ref{tab:strength}), and therefore galaxies must be sufficiently IR-bright. Nevertheless, as shown in Figure~\ref{fig9}e, our calculation indicates that we can expect to detect $\sim 30$ galaxies at $z=4-7$ in the PAH 3.3~$\mu$m feature, and even more if we consider gravitational-lensing effects (Egami et al. in prep.). Although the PAH 3.3~$\mu$m feature is relatively narrow, its profile is still resolved with SMI/LR, and thus even detecting a single feature may enable us to estimate the redshift (with the help of the comparably bright hydrogen recombination line Br$\alpha$ at 4.05~$\mu$m and possibly the sub-features at 3.4--3.6~$\mu$m and the H$_2$O ice feature at 3~$\mu$m.) Since SAFARI can measure rest-frame mid-IR PAH features while ALMA (Atacama Large Millimeter/submillimeter Array) can measure dust continuum emission, follow-up observations of those targets with SAFARI and ALMA are of particular importance to study organic matter chemistry and dust physics in the early universe.

In the context of a study of PAHs, we can expect a strong synergy
between JWST and SPICA; JWST can detect all near- and mid-IR features of
PAHs in the nearby universe, while SPICA can access those except the 17~$\mu$m feature only at $z>1$. JWST will be able to reveal detailed
physics and chemistry of PAHs in the nearby universe and to establish
the PAH features as spectral tools. Then, that knowledge can be
applied to studies of galaxies at high-$z$ with SPICA. Hence the roles of JWST and SPICA are complementary to each other in studying PAHs in the near and far universe. On the other hand, in terms of dusty AGN, we can expect a strong synergy between {\it Athena} (Advanced Telescope for High ENergy Astrophysics) and SPICA. The wide field-of-view of {\it Athena} enables efficient surveys of classical AGN, including both types 1 and 2. There is compelling evidence that the fraction of buried AGN increases with $L_{\rm IR}$ \citep{lee12}, and the SMI surveys are very sensitive to the buried AGN which {\it Athena} cannot easily detect due to heavy absorption \citep{gruppioni17}. The strategy of deep X-ray follow-up observations of SMI-selected AGN at high redshift will work very efficiently, because X-ray spectra above rest-frame 10 keV will provide crucial information on the poorly-studied nuclear environments of obscured AGN \citep{stern14,piconcelli15}.
Hence SPICA and {\it Athena} play roles complementary to each other in revealing contribution of both classical and buried AGN to $L_{\rm IR}$.

\section{Additional science}

It should be noted that future cosmological surveys with SMI/LR would simultaneously provide an unbiased, statistically significant view on nearby objects like foreground stars and galaxies as well. Among them, the low-resolution spectra of debris disks in main-sequence stars will be one of the most important by-products to take advantage of the high spectral survey speed of SMI/LR. We estimate below the number of the debris disks expected to be detected by the SMI/LR wide survey. 

First, based on the AKARI all-sky survey at 18~$\mu$m
\citep{ishihara10}, we estimate that a total number of $1.1\times10^4$
F, G, and K-type main-sequence stars would be detected at 20~$\mu$m by the SMI/LR wide survey, considering the improvement in the
sensitivity from AKARI to SPICA. Here we assume an isotropic
distribution of stars with 400 pc in the height of the Galactic disk in
the solar neighborhood \citep{siebert03}. AKARI detected debris disks of
luminosity levels $\sim 1\times 10^3$ times higher than that of our zodiacal cloud \citep[L$_{\rm Zodi} \simeq 1 \times 10^{-7}$~L$_{\odot}$;][]{nesovorny10} in an unbiased manner, and revealed that $\sim 10$\% of the stars possess debris disks at that luminosity threshold \citep{ishihara17}. Then we estimate what fraction of the main-sequence stars detected in the SMI survey possess detectable debris disks, assuming a luminosity function of debris disks, which is unknown for faint disks and thus could be determined by SPICA. 

\begin{figure}[htbp]
\begin{center}
  \includegraphics[width=\columnwidth]{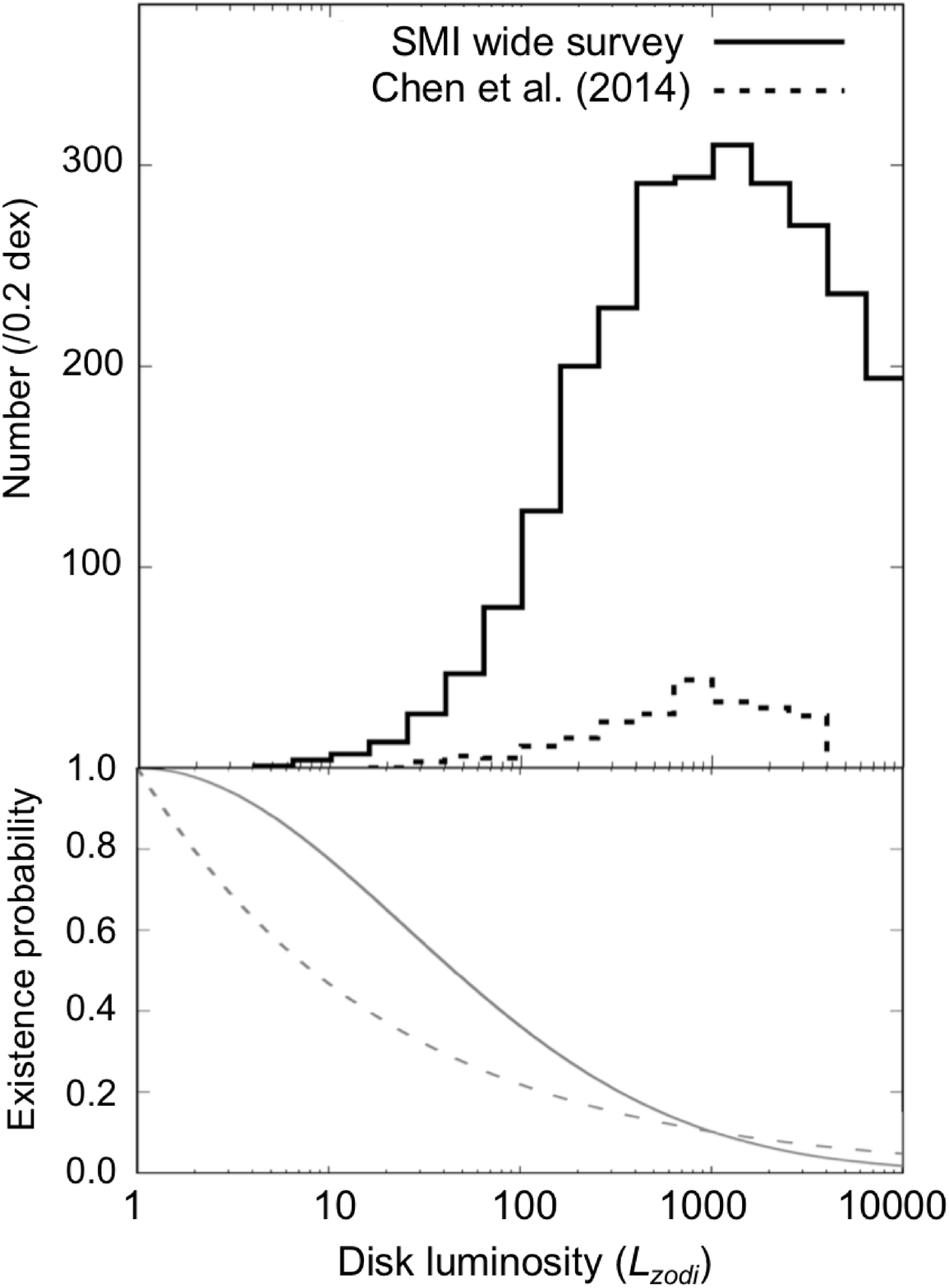}
    \caption{Numbers of the debris disks of F, G, and K-type main-sequence stars expected to be detected by the SMI/LR wide survey, which are estimated based on the result of the AKARI all-sky survey as a function of the disk luminosity. The {\it Spitzer} summary result is also shown for comparison \citep{chen14}. The lower panel shows the assumed existence probability curves of debris disks as a function of the disk luminosity, the solid one of which is used to obtain the result of the upper panel. The dashed curve is also considered to estimate the uncertainty of the result.}
\label{fig10}
\end{center}
\end{figure}

Estimating the limiting flux density of dust emission from debris disks
is not straightforward, because we have to consider the underlying
photospheric continua of the central stars. As a rough estimation, we
adopt the limiting flux density of 200~$\mu$Jy at 20~$\mu$m, which is 10
times worse than the 5$\sigma$ continuum sensitivity of SMI/LR in a low
background; this flux density corresponds to the luminosity of a
debris disk with L$_{\rm Zodi}$ at a distance of 10 pc for a 200 K
blackbody continuum emission. Figure~\ref{fig10} shows the numbers of the
debris disks estimated with the above limiting flux density under the
assumption that the (cumulative) existence probability of debris disks
with $>1\times 10^3$~L$_{\rm Zodi}$ and $>$1~L$_{\rm Zodi}$ are 10\%
(based on the AKARI results), and 100\%, respectively. We
interpolated the existence probability function between 1~L$_{\rm Zodi}$
and $1\times 10^3$~L$_{\rm Zodi}$ by two types of curves as shown in the
lower panel of Figure~\ref{fig10}, and counted the numbers of debris
disks in the luminosity range of $1$~L$_{\rm Zodi}$ to $1\times 10^4$~L$_{\rm Zodi}$. As a result, the total number is in the 1800--2600 range, depending on the types of the existence probability curves. The probability of signal blending with galaxies is very low; the number of galaxies expected to be detected with the above limiting flux density is calculated to be about $2\times 10^{-4}$ per beam ($3.7''$) of SMI/LR in the wide survey. 

Our simple calculation suggests that we can detect faint debris disks of 5--10~L$_{\rm Zodi}$ levels in principle. It is a significant step forward to statistically understanding the evolution of debris disks towards our zodiacal cloud analogues, with fainter and presumably more common disks than those in the previous studies \citep[e.g.,][]{chen14}. However, in order to detect such faint debris disks, we must determine the underlying stellar continuum with 0.1--1\% calibration accuracy and stability, which is in practice rather difficult. One possibility is to use different spectral shapes between the Rayleigh-Jeans stellar continuum and thermal dust emission, especially dust bands if present \citep[e.g., silicate features; ][]{fujiwara10, olofsson12}. Hence the SMI/LR wide survey could characterize the properties of many debris disks and potentially detect faint debris disks similar to our zodiacal cloud.

\section{Summary}
We have evaluated the capability of low-resolution mid-IR spectroscopic
surveys of galaxies with SMI onboard SPICA. For instance, a wide
survey of 10 deg$^2$ area in 600 hours would provide $\sim 5\times10^4$
PAH spectra from galaxies at $z > 1$, would detect more than
$2\times10^5$ dusty AGN at 34~$\mu$m with the slit viewer, and at the
same time, is expected to obtain more than $1\times10^4$ spectra
from main-sequence stars of F, G, and K types in the foreground and at
least $1\times10^3$ debris disks among them. Thus the SMI/LR-CAM
surveys are capable to efficiently provide us with unprecedented large
spectral and photometric samples that would cover very nearby
planet-forming stars to distant star-forming galaxies and active
galactic nuclei especially in the unexplored 30$-$40~$\mu$m wavelength
regime. These samples would be crucial as follow-up candidates to be
further studied with SPICA on such major science topics as described in
a series of the relevant papers in this volume \citep{spinoglio17,
gonzalez17, fernandez17, gruppioni17, vandertak17, egami17}. 

\begin{acknowledgements}
This paper is dedicated to the memory of Bruce Swinyard, who unfortunately passed away on 22 May 2015 at the age of 52. He initiated the SPICA project in Europe as first European PI of SPICA and first design lead of SAFARI. 

We thank all the members of SPICA Science Working Group and the SMI consortium for their continuous discussions on science case and requirements for SMI. We are especially grateful to the board members of SPICA Science Case International Preview (Michael Rowan-Robinson, Martin Bureau, David Elbaz, Peter Barthel, Anthony Peter Jones, Martin Harwit, George Helou, Kazuhisa Mitsuda) and JAXA's SPICA International Science Advisory Board (Philippe Andre, Andrew Blain, Michael Barlow, David Elbaz, Yuri Aikawa, Ewine van Dishoeck, Reinhard Genzel, George Helou, Roberto Maiolino, Margaret Meixner, Tsutomu Takeuchi) for giving us invaluable comments and advice. The optical/mechanical designing activities of SMI to fulfill the science requirements are funded by JAXA within the framework of the SPICA preproject in Phase A1.
\end{acknowledgements}

% UNCOMMENT THE LINES BELOW IF YOU WISH TO USE BIBTEX

\nocite*{}
\bibliographystyle{pasa-mnras}
\bibliography{1r_lamboo_notes}

\end{document}